\documentclass[10pt,a4paper,twocolumn]{IEEEtran} 
\normalsize
\usepackage[colorlinks,
linkcolor=black,%red,
anchorcolor=black,%blue,
citecolor=black,%green
urlcolor=black,
]{hyperref}
\usepackage[hyphenbreaks]{breakurl}
\usepackage{diagbox}
\usepackage[english]{babel}
\usepackage{latexsym}
\usepackage{varioref}
\usepackage{mathrsfs}
\usepackage{array}
\usepackage{hhline}
\usepackage{dcolumn}
\usepackage{tabularx}
\usepackage{algorithm}
\usepackage{algorithmic}
\usepackage{calc}
\usepackage{epsfig}
\usepackage[usenames]{color}
\usepackage{amsmath,amssymb,bigdelim}
\usepackage{t1enc}
\usepackage{pstricks,pst-node,psfrag,pst-plot}
\usepackage{graphicx,parskip,amsbsy}
\usepackage{boxedminipage}
\usepackage{indentfirst}
\usepackage{booktabs}
\usepackage[noadjust]{cite}
\usepackage{caption}
\usepackage{multirow}
\usepackage{subfigure}
\usepackage[usenames]{color}
\usepackage{clrscode}
\usepackage{url}
\usepackage{setspace}
\usepackage{pstool}
\usepackage{fancybox}
\usepackage[shortcuts,acronym]{glossaries}
\newacronym{5G}{5G}{fifth-generation}
\newacronym{MIMO}{MIMO}{multiple-input-multiple-output}
\newacronym{SISO}{SISO}{single-input-single-output}
\newacronym{BS}{BS}{base station}
\newacronym{UE}{UE}{user equipment}
\newacronym{PAS}{PAS}{power angular spectrum}
\newacronym{SMPAC}{SMPAC}{sectored MPAC}
\newacronym{SMPAC in abstract}{SMPAC}{sectored multiprobe anechoic chamber}
\newacronym{MPAC}{MPAC}{multiprobe anechoic chamber}
\newacronym{RTS}{RTS}{radiated two stage}
\newacronym{RC}{RC}{reverberation chamber}
\newacronym{HPBW}{HPBW}{half-power-beamwidth}
\newacronym{UCA}{UCA}{uniform circular array}

\newacronym{RF}{RF}{radio frequency}
\newacronym{OTA}{OTA}{over-the-air}
\newacronym{SNR}{SNR}{signal-to-noise-ratio}
\newacronym{mmWave}{mmWave}{millimeter-Wave}
\newacronym{RRM}{RRM}{radio resource management}
\newacronym{PWS}{PWS}{plane wave synthesis}
\newacronym{PFS}{PFS}{prefaded signal synthesis}
\newacronym{CSI}{CSI}{channel state information}
\newacronym{CDL}{CDL}{clustered-delay-line}
\newacronym{DUT}{DUT}{device under test}

\newacronym{CST}{CST}{computer simulation technology}

\newacronym{VNA}{VNA}{vector network analyzer}
\newacronym{CTF}{CTF}{channel transfer function}
\newacronym{MPC}{MPC}{multipath component}
\newacronym{LoS}{LoS}{line-of-sight}
\newacronym{NLoS}{NLoS}{non-line-of-sight}
\newacronym{3GPP}{3GPP}{3rd Generation Partnership Project}

\newacronym{PXI}{PXI}{PCI eXtensions for Instrumentation}
\newacronym{mMIMO}{mMIMO}{massive multiple-input-multiple-output}
\newacronym{UAV}{UAV}{unmanned aerial vehicle}
\newacronym{SDR}{SDR}{Software Defined Radio}
\newacronym{CIR}{CIR}{channel impulse response}
\newacronym{SINR}{SINR}{signal-to-interference-plus-noise ratio}
\newacronym{USRP}{USRP}{Universal Software Radio Peripheral }
\newacronym{APDP}{APDP}{averaged power delay profile}
\newacronym{PDP}{PDP}{power delay profile}
\newacronym{SAGE}{SAGE}{space-alternating generalized expectation-maximization}
\newacronym{HRPE}{HRPE}{high-resolution-parameter-estimation}
\newacronym{LTE}{LTE}{Long Term Evolution}
\newacronym{FDD}{FDD}{frequency division duplex}
\newacronym{EM}{EM}{expectation-maximization}
\newacronym{AIC}{AIC}{Akaike information criterion}
\newacronym{KPM}{KPM}{K-Power-Means}
\newacronym{MCD}{MCD}{multipath component distance}
\newacronym{GMM}{GMM}{Gaussian-mixture-model}
\newacronym{CH}{CH}{Cali\~nski-Harabasz}
\newacronym{DB}{DB}{Davies-Bouldin}
\newacronym{RMS}{RMS}{root-mean-square}
\newacronym{A2G}{A2G}{air-to-ground}
\newacronym{C2}{C2}{command and control}
\newacronym{AoA}{AoA}{azimuth of arrival}

\makeatletter
\def\url@leostyle{%
  \@ifundefined{selectfont}{\def\UrlFont{\sf}}{\def\UrlFont{\small\ttfamily}}}
\makeatother
\newcommand\scalemath[2]{\scalebox{#1}{\mbox{\ensuremath{\displaystyle #2}}}} % resize equations
\newcommand{\bs}{\boldsymbol}

\newcommand{\bsTheta}{{\bs \Theta}}

\newcommand{\rem}[1]{}
\newcommand{\addnewfr}[1]{{\color{black} #1}}
\renewcommand{\baselinestretch}{1}%0.99
\providecommand{\texlivekeywords}[1]{\textbf{\textit{Index terms---}} #1}
\providecommand{\fref}[1]{Fig.\,\ref{#1}}
\providecommand{\frefp}[1]{Figs.\,\ref{#1}}
\providecommand{\sref}[1]{Sect.\,\ref{#1}}
\begin{document}

	\pagestyle{plain}
	\title{Empirical Low-Altitude {Air-to-Ground} Spatial Channel {Characterization} for Cellular Networks Connectivity}

  \author{Xuesong Cai, Tomasz Izydorczyk, Jos\'e Rodr\'iguez-Pi{\~n}eiro,
  Istv\'an Z. Kov\'acs\\ Jeroen Wigard, Fernando M. L. Tavares and Preben E. Mogensen
	
		\thanks{
%This work has been submitted to the IEEE (J-SAC) for possible publication. Copyright may be transferred without notice, after which this version may no longer be accessible.

		X. Cai is with the Department of Electronic Systems, Aalborg University, 9220 Aalborg, Denmark (email: xuesong.cai@ieee.org).

T. Izydorczyk is with Nokia Mobile Networks, 53-611 Wroclaw, Poland (email: tomasz.izydorczyk@nokia.com).

J. Rodr\'iguez-Pi{\~n}eiro is with the College of Electronics and Information Engineering, Tongji University, 201804 Shanghai, China (e-mail: j.rpineiro@tongji.edu.cn).

I. Z. Kov\'acs and J. Wigard are with Nokia Bell Labs, 9220 Aalborg, Denmark (email: istvan.kovacs@nokia-bell-labs.com; jeroen.wigard@nokia-bell-labs.com). %, 9220 Aalborg, Denmark.

%X. Yin is with the College of Electronics and Information Engineering and the National Computer and Information Technology Practical Education Demonstration Center, Tongji University, 201804 Shanghai, China (e-mail: yinxuefeng@tongji.edu.cn).

F. M. L. Tavares is with QIAGEN, 8000 Arhus, Denmark. (email: fmltavares@gmail.com)

P. E. Mogensen is with the Department of Electronic Systems, Aalborg University, and Nokia Bell Labs, 9220 Aalborg, Denmark (e-mail: pm@es.aau.dk).

% Juyul Lee is with the Telecomm. Media Research Lab, Electronics \& Telecomm. Research Institute (ETRI), Daejeon, Korea (e-mail: juyul@etri.re.kr).

% This work was partially supported by the National Natural Science Foundation of China (NSFC) under Grants 61850410529 and 61971313.

   }
%\vspace{-2.1cm}
	}

\markboth{IEEE Transactions on Communications}%
{Submitted paper}

\maketitle \thispagestyle{plain}

\begin{abstract}
%Enabling ubiquitous air-space-ground communications is an important vision in the \ac{5G} and beyond. For this purpose,
Cellular-connected \acfp{UAV} %As part of the vision in the \acf{5G} networks and beyond enabling ubiquitous air-space-ground communications, cellular-connected \acfp{UAV}
have recently attracted a surge of interests in both academia and industry. Understanding the \acf{A2G} propagation channels is essential to enable reliable and/or high-throughput communications for \acsp{UAV} and protect the ground \acfp{UE}. In this contribution, a recently conducted measurement campaign for the \acs{A2G} channels is introduced. A \acf{UCA} with 16 antenna elements was employed to collect the downlink signals of two different \acf{LTE} networks, at the heights of 0-40\,m in three different, namely rural, urban and industrial scenarios. The \acfp{CIR} have been extracted from the received data, and the spatial, including angular, parameters of the multipath components in individual channels were estimated according to a \acf{HRPE} principle. Based on the \acs{HRPE} results, clusters of multipath components were further identified. Finally, comprehensive spatial channel characteristics were investigated in the composite and cluster levels at different heights in the three scenarios. %The obtained results of the \acs{A2G} spatial channels are important for the vision of \acs{5G} and beyond.
%, with high fidelity to that experienced by the real cellular-connected \acs{UAV}-\acsp{UE},

% As a part of the vision in the \ac{5G} networks and beyond enabling ubiquitous air, space and ground communications, cellular-connected \acp{UAV} have recently attracted a surge of interests in both academia and industry, with a big commercial market foreseen in the near future. Understanding the \ac{A2G} channels is essential to propose advanced techniques and solutions to enable reliable and/or high-throughput communications for \acp{UAV} as well as protect the ground \acp{UE}. In this contribution, a recently conducted measurement campaign for the \ac{A2G} channels is introduced. A real \ac{UCA} consisting of 16 antenna elements was exploited to collect the downlink signals of \ac{LTE} networks, at the heights from 0 to 40\,m in three different, i.e. rural, urban and industrial, scenarios. The \acp{CIR} are extracted from the received downlink data, and the spatial/angular parameters of the multiple path components in individual channels are estimated according to a \ac{HRPE} principle. Based on the \ac{HRPE} results, clusters of the multipath components are further identified. Finally, comprehensive spatial channel characteristics extracted in the composite and cluster levels at different heights in the three scenarios are elaborated. The obtained results of the \ac{A2G} spatial channels, with high fidelity to that experienced by the real cellular-connected \ac{UAV}-\acp{UE}, are invaluable for the vision of the \ac{5G} and beyond networks.

\end{abstract}
\texlivekeywords{UAV, cellular networks, air-to-ground, spatial channels, angular characteristics, and clusters.}
\IEEEpeerreviewmaketitle

% !TeX spellcheck = en_US
\section{Introduction}

Recently, \acp{UAV} have been shifting their use from purely military operations to a more general-purpose scope with rapidly increasing popularity, due to the continuous reduction of cost, size, weight and consumption. A huge market is foreseen with many new civilian and commercial applications such as forest monitoring, goods delivery or search and rescue in hostile environments \cite{hayat2016survey,kumar2018unmanned,menouar2017uav}.%\cite{kumar2018unmanned}. 
%\cite{hayat2016survey,kumar2018unmanned,menouar2017uav}.
%{\red \cite{kumar2018unmanned} \textit{(some references removed here)}}.
Temporary network access provided by \acp{UAV} as movable aerial \acp{BS} in emergency situations or saturated communication environments has become a key scenario addressed for the \ac{5G} and beyond communication systems \cite{zeng2016wireless}. Unlike terrestrial \acp{BS}, for which the power supply is in general not an issue, \acp{UAV} have limited power constraints for the signal transmission, processing and the \ac{UAV} propulsion. This generated an increasing research interest, e.g., on the \ac{UAV} flight trajectories design, that jointly optimizes the throughput, latency and energy consumption of the network \cite{wu2019fundamental,wu2018uav,zeng2017energy}. Meanwhile, most of the civilian-used \acp{UAV} still rely on direct point-to-point communications with their ground pilots. This usually limits their usage to the visual \ac{LoS} region. Further, the lack of coordination among \acp{UAV} cannot guarantee the network scaling with increasing number of
\acp{UAV}. Moreover, other regulatory issues \cite{stocker2017review}, e.g. forbidding unregistered \acp{UAV} or the flights near to the airports, have also been widely concerned. Therefore, the so-called cellular-connected \acp{UAV} \cite{8470897,zeng2019accessing}, i.e. using the existing cellular infrastructure to provide reliable \ac{C2} communications and high-throughput payload communications such as high-definition videos for \ac{UAV}-\acp{UE}, have attracted a surge of interest in both academia and industry \cite{3gppuasuav,8581827} in \ac{5G} and beyond communication systems. Although preliminary field investigations \cite{nguyen2018how,amorim2020enabling,9082692} have shown the potential of cellular networks in providing reliable \ac{C2} and high-throughput payload communications, there are still challenges that may lead to the system bottlenecks. For example, cellular \ac{BS} antennas are usually targeted to serve ground users and hence down-tilted; \ac{UAV}-\acp{UE} in the sky may cause severe interferences \cite{ArtigoInterferenciaEnUAVs_2019,8369158} among them and to the ground \acp{UE} in both up- and downlinks, which limits the spectrum efficiency significantly. Several investigations dealing with the interferences through power control, network coordination, resource allocation, beam usage, etc. to increase the system capacity and protect the ground \acp{UE} have been conducted \cite{cai2020centralized,9082692,9099899,8763928,UAVResourceAllo}. In order to properly design and optimize the proposed techniques, realistic channel models for the \ac{A2G} propagation channels in different environments are essential.

Due to the regulatory limitations in the flight height of \acp{UAV} in most countries \cite{european2013roadmap}, low-altitude \ac{A2G} propagation channels for small-sized \acp{UAV} have recently attracted considerable attention. {Simulation-based approaches have been employed. For example, ray-tracing tools were utilized in \cite{8288376,8244753,8674482} with path loss, channel dispersions, etc. analyzed. In \cite{9211744,8594724,8937764,8633886} and references therein, geometry-based stochastic models consisting of different stochastic properties were investigated by assuming that scatterers distribute with certain patterns.}
%{\red Simulation-based studies have been employed for proposing \ac{UAV} \ac{A2G} channel models such as in the ray-tracing based works and  the two-ray one in \cite{8288376} or the path loss and delay spread in \cite{8244753} and \cite{8674482}. \cite{9211744,8594724,8633886}
{Nevertheless, measurement-based works may be preferred since some of the assumptions for simulation-based works are not proved to be realistic. For example, our previous works \cite{ArtigoCaracteristicasCanleComunicacionsUAV_2017,8576578} have shown that the two-ray assumption \cite{8288376} may not be appropriate for low heights.}
{Narrowband} channel characteristics such as path loss {obtained by means of actual measurements can be found in e.g.} \cite{hourani2018modeling}. {In this work, the authors characterize the \ac{A2G} path loss as an excess factor with respect to a terrestrial user path loss}. {Shadowing was also considered in \cite{amorim2017radio}, which is also limited to narrowband case.} {Wideband channel characterization based on measurements was in turn considered in \cite{khawaja2016uwb}, although the flight height is limited to $16$\,m}.
%Propagation channel characteristics such as path loss \cite{hourani2018modeling,wang2017path,amorim2017radio}, shadowing \cite{amorim2017radio} and small-scale fading \cite{khawaja2016uwb} were studied based on measurements. 
In our previous works \cite{ArtigoCaracteristicasCanleComunicacionsUAV_2017,8576578}, we have proposed several stochastic channel models for different flight heights, scenarios and \ac{BS} antennas configurations. These channel models include characterizations of the path loss, shadow fading, delay spread, Doppler frequency spread and K-factor. The obtained results, especially those in
\cite{ArtigoCaracteristicasCanleComunicacionsUAV_2017}{, which were also corroborated in \cite{ArtigoCaracteristicasCanleComunicacionsUAV_ParteTraxectorias_2017}}, exhibited some non-intuitive effects. For example, it was shown that increasing the flight height could lead to a decrease of the K-factor due to strong reflections from tall buildings, which are blocked by objects close to the \ac{UAV} when the flight height is lower. This indicates that simply assuming the \ac{A2G} channel with a high flight height to be \ac{LoS} dominant {or follow a simple two-ray model}, which is valid for traditional large aircrafts with very high flight altitudes, e.g. above 1\,km, may be non-realistic for the low-altitude (below 120\,m) small-sized \acp{UAV}. %{\red \textit{(some reference removed here)}}
%Besides the above mentioned investigations, there are still many other works regarding the low-altitude \ac{A2G} channels \cite{8787874}.
{Many more measurement-based works can also be found in \cite{8787874,8576578} and references therein.}\footnote{{It is worth noting that although D. Matolak \textit{et al.} provided a rich set of measurement-based channel models in different environments (e.g., see \cite{7486380}), they are suited for small aircrafts flying at high heights (more than $500$\,m) and much larger distances between the BS and the aircraft (several kilometers), being not suitable for small-sized \acp{UAV}.}}
{However}, most of the measurement-based investigations only focused on the \ac{SISO} and temporal channel characteristics. In \cite{8928089,izydorczyk2019angular}, very preliminary composite angular characteristics were reported. There seems, to the best of the authors' knowledge, to be no comprehensive measurement-based investigation on the spatial characteristics of the low-altitude \ac{A2G} channels. One probable reason is the
%is the difficulty in establishing the measurement systems considering the limited load capacity of small-sized \acp{UAV}.
difficulty in designing powerful \ac{MIMO} channel measurement systems with acceptably low weights that can be installed on a small-sized \ac{UAV} with limited load capacity. However, the spatial characteristics of the low-altitude \ac{A2G} channels are essential to evaluate or verify various solutions proposed, e.g. the \ac{mMIMO} or beamforming based techniques
\cite{9082692,garcia2019essential,geraci2018understanding}
%\cite{9082692} {\red \textit{(some references removed here)}}
with the potential to mitigate interferences and increase the network capacity.
% such as the exploitation of \ac{mMIMO} for spatial multiplexing \cite{garcia2019essential,geraci2018understanding} or beam selection for spatial diversity \cite{9082692}.

In order to fill the research gap, in this paper we substantially extend our previous analysis \cite{izydorczyk2019angular} to properly characterize the low-altitude \ac{A2G} spatial channels in multiple scenarios. A channel sounder based on a \ac{UCA} featuring 16 elements was exploited to sound the \ac{A2G} channels from 0 to 40\,m in three typical scenarios, namely rural, industrial and urban, and a complete set of spatial characteristics of the channels on both the composite and cluster levels is proposed. Up to the authors' knowledge, this is the first contribution of this kind for the low-altitude \ac{A2G} communication channels. The empirically obtained results interestingly show e.g. that though in general the \ac{A2G} channels become less complicated (e.g. with smaller spreads and less clusters) when the height is increased, it is possible that an increase of the height in a complex (urban or industrial) scenario can lead to more complicated channels consisting of several well-separated spatial clusters with similar powers that can be potentially exploited to increase the system capacity. This work not only confirms our previous findings in the temporal domain reported in e.g. \cite{ArtigoCaracteristicasCanleComunicacionsUAV_2017,ArtigoCaracteristicasCanleComunicacionsUAV_ParteTraxectorias_2017}, but also characterizes the spatial characteristics comprehensively considering the angular domain as well as cluster level behaviours. The main contributions and novelties which lead to a much deeper understanding of the propagation mechanisms for low-altitude \ac{A2G} communications in this paper are summarized as follows:
\begin{itemize}
  \item A \ac{SDR}-based channel sounder featuring a \ac{UCA} with 16 elements was exploited for channel sounding, where the real-time downlink signals of the commercial \ac{LTE} networks were captured at the heights from 0 to 40\,m in the rural, industrial and urban scenarios. The measurement methodology makes it possible that the \acp{CIR} extracted from the measured data are with high fidelity to those experienced by \ac{UAV}-\acp{UE} in cellular networks.

  \item   High-resolution channel parameters of multipath components are estimated using a proposed \ac{HRPE} principle with decent calibrations. This allows us to further identify clusters in individual channels.

  \item Based on the cluster identification results, spatial characteristics of the \ac{A2G} channels on the composite and cluster levels are investigated thoroughly, which makes it realizable to reproduce the measured channels in stochastic sense as specified in the standard modelling structure \cite{3GPP38901}.
\end{itemize}

The rest of the paper is organized as follows. Sect.\,\ref{section:setup_and_environment} elaborates the measurement methodology including hardware design and scenarios. Sect.\,\ref{sect:cirextraction_and_apdp} discusses the \ac{CIR} extraction and presents some typical \acp{APDP} in different scenarios.
In Sect.\,\ref{sect:data_processing}, the channel parameter estimation and cluster identification algorithms applied for the measured channels are elaborated. Based on Sect.\,\ref{sect:data_processing}, comprehensive characteristics of the spatial \ac{A2G} channels in composite and cluster levels are investigated and summarized in Sect.\,\ref{sect:channel_models}. Finally, conclusive remarks are included in Sect.\,\ref{section:conclusion}.

\section{Measurement methodology}\label{section:setup_and_environment}
In this section, the measurement testbed is presented followed by the description of the measurement campaigns. Since both were already thoroughly described in~\cite{8894135}, this section focuses only on the main aspects that are important to understand the performed studies.
\subsection{Measurement design}\label{section:design}
Spatial channel characteristics are one of the key factors influencing the performance of \ac{mMIMO} systems. In order to best evaluate the potential of \ac{mMIMO} in a context of \ac{UAV} communication, these characteristics need to be described focusing on their height dependency. %Different channel parameters should be observed at very low altitudes when a \ac{UAV} flies between buildings, then at the heights right above the rooftops, and finally yet another results are expected when \acp{UAV} fly at high altitudes.
Measurements using live cellular networks, as proposed in this work, offer the possibility to study channel properties in the real environment that will be observed by the flying \acp{UAV}. They also offer a simple yet effective solution to collect, in the same measurement snapshot, the multiple signals coming from different serving cells that may potentially be decoded.

\subsection{Measurement hardware}\label{section:hardware}
\begin{figure}
\centering
\includegraphics[width=0.5\textwidth]{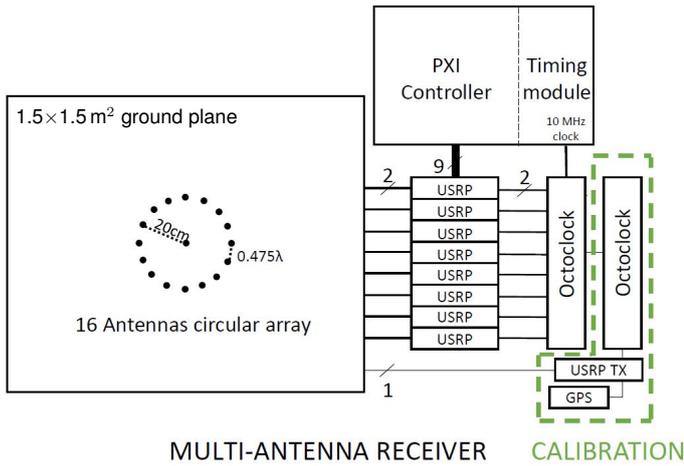}{
\psfrag{2.25}[l][l][0.75]{\textsf{1.5$\times$1.5\,m$^2$ ground plane}}
}
\caption{Measurement system schematic \cite{8894135}. 16 antennas composed a UCA, and 1 antenna was located in the center of the UCA for calibration purpose. 9 \acp{USRP} were utilized to provide the 17 \ac{RF} chains.}
\label{fig:schemat}
\end{figure}

\begin{figure}
\centering
\subfigure[]{\includegraphics[width=0.3\textwidth]{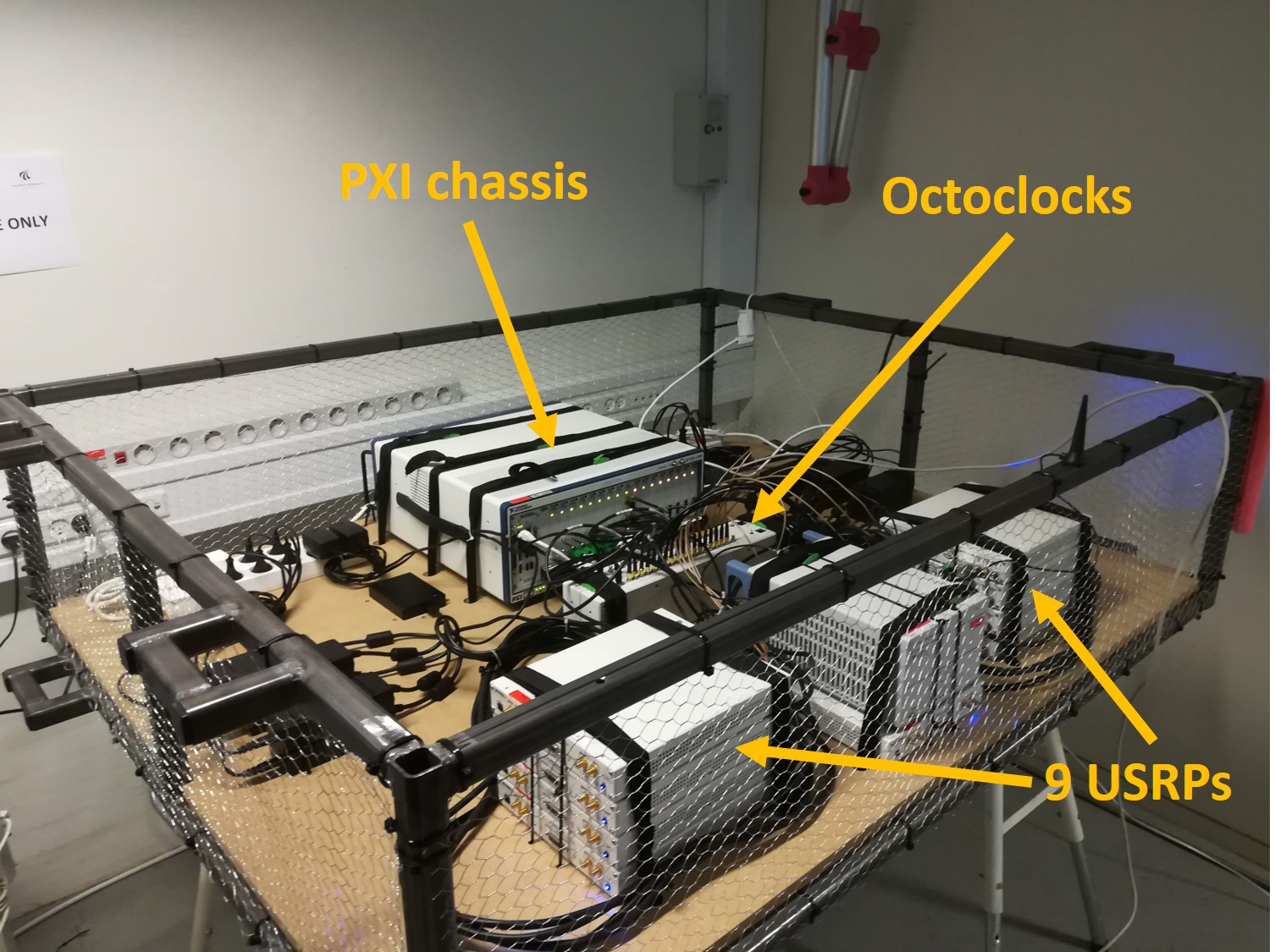}}
\subfigure[]{	{
    {\includegraphics[width=0.3\textwidth]{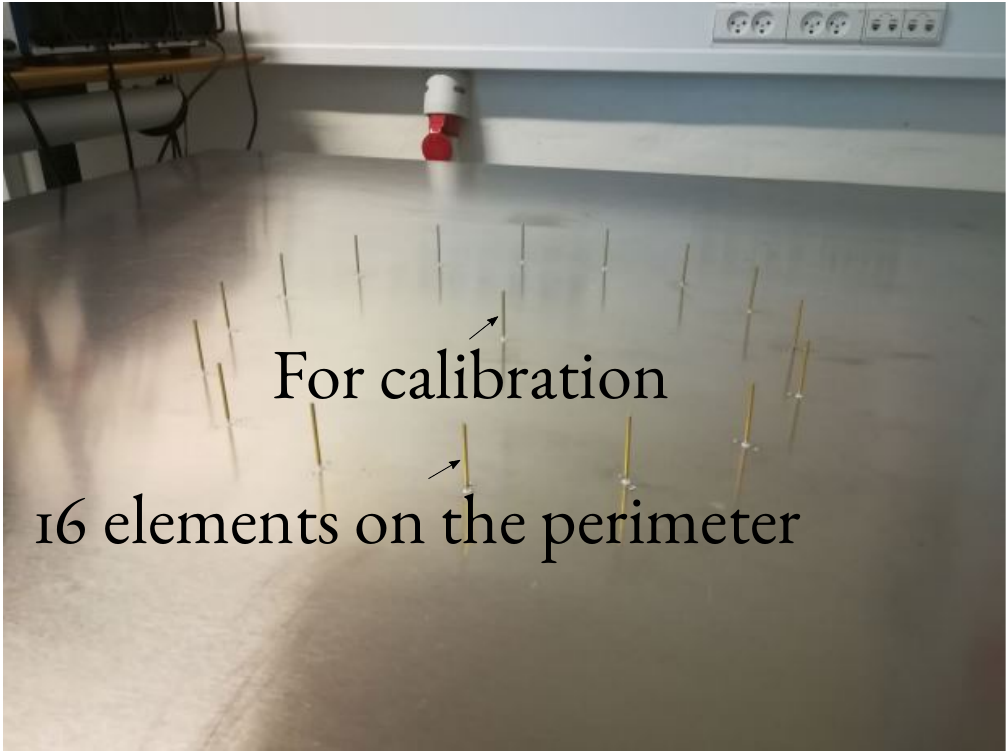}}}{
\psfrag{A}[c][c][0.9]{16 elements on the perimeter}
\psfrag{B}[c][c][0.9]{For calibration}
}}
\caption{Measurement equipment. (a) \ac{SDR} test-bed. (b) The \ac{UCA}. The antenna in the center is for calibration of \ac{USRP} boards.}
\label{fig:equipment}
\end{figure}

The measurement setup was built based on nine \ac{USRP} boards. They are used as digital \ac{RF} chains to connect sixteen antennas forming a \ac{UCA}. Further, \acp{USRP} are connected to the \ac{PXI} chassis, used to record and store raw recorded complex data samples, without any real time processing. Besides, two octoclocks are used to provide time and frequency synchronization among all \ac{USRP} boards. \fref{fig:schemat} presents the schematic of the testbed, meanwhile in \fref{fig:equipment}(a) the actual implementation inside the steel cage before installation of an antenna array is presented.

%\subsubsection{Antenna array design}
\textit{1) Antenna array design}:
The sixteen-antennas \ac{UCA}, as illustrated in \fref{fig:equipment}(b), was designed to operate at \ac{LTE} band~3 (1.8~GHz frequency band), where two \ac{LTE} networks belonging to the different Danish telecom operators are located. Antenna elements, realized as simple monopoles are distributed with around 0.475\,$\lambda$ spacing, corresponding to a circular radius of 20~cm and are manufactured on a large 1.5$\times$1.5\,m$^2$ aluminum ground plane installed on top of the presented steel cage. The final antenna radiation pattern was simulated using \ac{CST} studio and can be well approximated by an omni-directional pattern with approximately 10\textsuperscript{o} up-tilt due to the finite size of the ground plane.

%\subsubsection*{Calibration of USRP boards}
\textit{2) Calibration of USRP boards}:
Spatial channel characterization requires all \ac{RF} chains to record the data with time, frequency and phase synchronization. Octoclocks are used to provide sufficient time and frequency alignment. However, testbeds containing multiple \ac{USRP} boards suffer from the inherited lack of phase synchronization between them. Although boards are phase coherent, each of them starts with the random phase offset. Therefore, external, self-designed calibration to compensate this offset is required.
The designed testbed, uses the additional antenna and a self-transmitted calibration signal for phase compensation. By transmitting a sine wave with a low power (-50~dBm) and a frequency near to the \ac{LTE} band using the antenna located in the center of the array, each of the antennas forming the \ac{UCA} are expected to receive the signal with the same time delay and the same phase. Assuming the received phase of one antenna as a reference, the phases of the remaining fifteen antennas can be compensated, allowing phase synchronization with less than 1\textsuperscript{o} error.

\subsection{Measurement campaign}\label{section:campaign}

\begin{figure}
\centering

\subfigure[]{\includegraphics[width=0.4\textwidth]{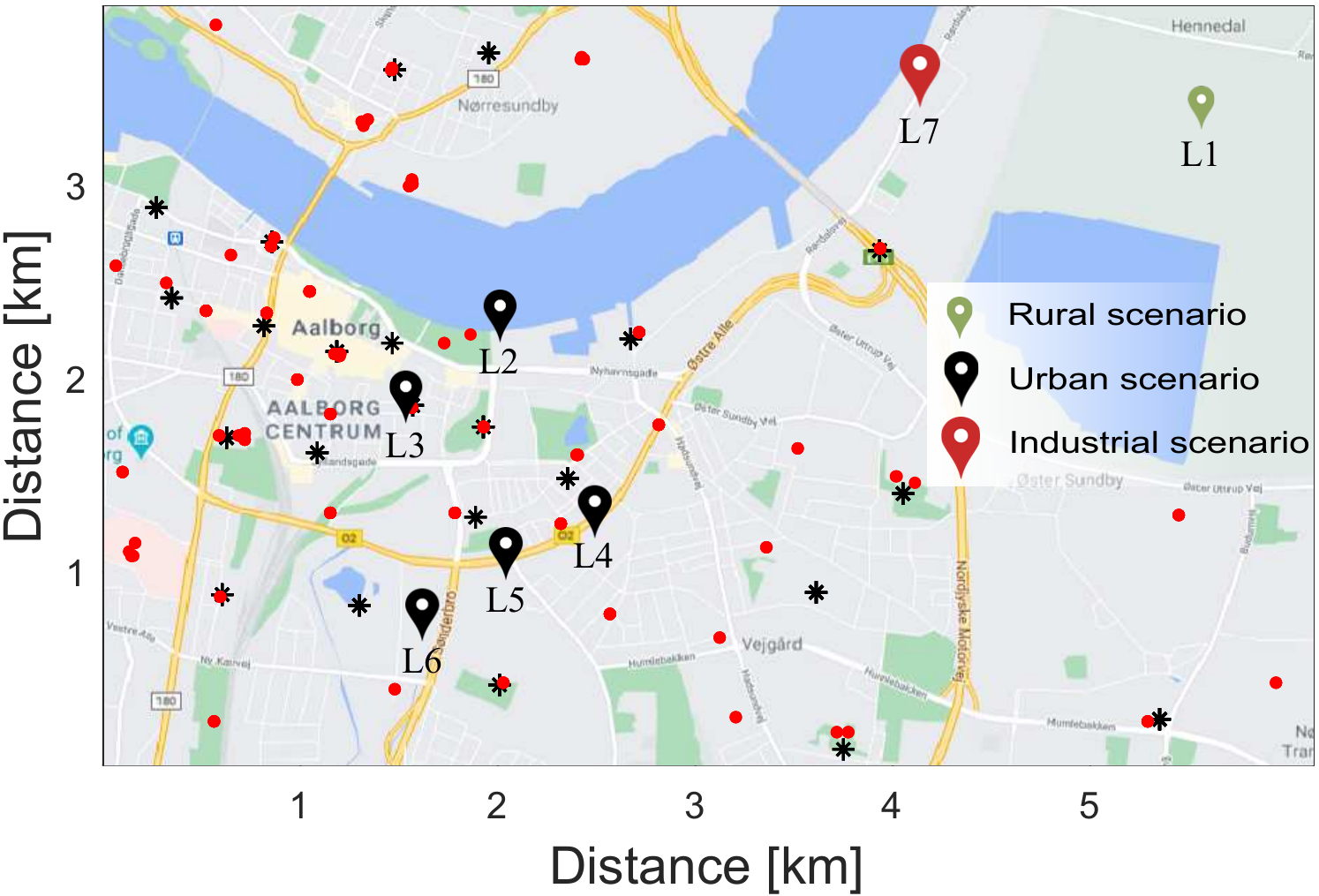}{
\psfrag{1}[c][c][0.7]{L1}
\psfrag{2}[c][c][0.7]{L2}
\psfrag{3}[c][c][0.7]{L3}
\psfrag{4}[c][c][0.7]{L4}
\psfrag{5}[c][c][0.7]{L5}
\psfrag{6}[c][c][0.7]{L6}
\psfrag{7}[c][c][0.7]{L7}
\psfrag{I}[l][l][0.75]{Industrial scenario}
\psfrag{U}[l][l][0.75]{Urban scenario}
\psfrag{R}[l][l][0.75]{Rural scenario}
}\label{fig:scenaris_a}} % replace with the google map figure

\subfigure[]{\includegraphics[width=0.3\textwidth]{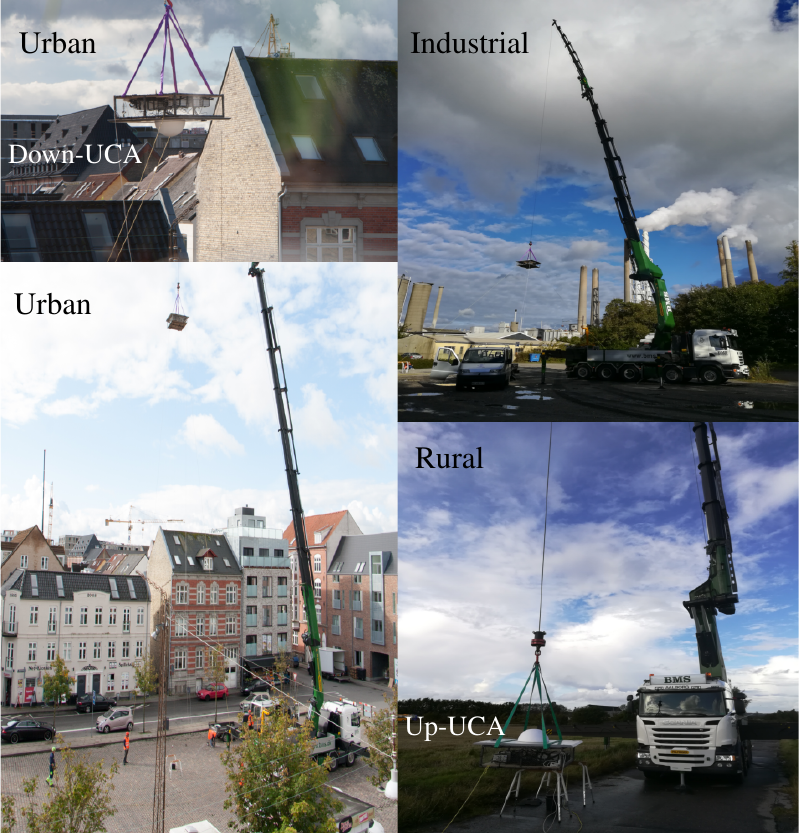}{
\psfrag{Urban}[l][l][0.9]{Urban}
\psfrag{I}[l][l][0.9]{Industrial}
\psfrag{R}[l][l][0.9]{Rural}
\psfrag{Up}[l][l][0.8]{{\white Up-UCA}}
\psfrag{Down}[l][l][0.8]{{\white Down-UCA}}
}\label{fig:scenaris_b}}
\caption{Measurement scenarios. (a) Measurement locations in the map. {The red dots and black asterisks indicate the {BS} locations of two network operators, respectively.} (b) Assembled setup lifted in different scenarios.} % urban scenario with antennas located on the top of the cage (big picture) and at the bottom (top left corner) {\red Hi Tomasz, you can send me some figure in industrial and rural areas (and the current two in this figure), I can combine selected figures for the paper.}}
\label{fig:photomeas}
\end{figure}

As illustrated in \fref{fig:photomeas}, the measurement campaign was conducted in seven different locations located in a vicinity of Aalborg, Denmark. It comprises three different measurement scenarios as described below:\footnote{\addnewfr{The scenarios were classified accorinding to the geographical locations and the observed similarity of channel characteristics. Specifically, locations\,2-6 were in Aalborg downtown and grouped as one scenario. Location\,7 and 1 were both in the countryside. However, due to the factory buildings in location\,7, we found that the channel characteristics observed in location\,7 were very different from that observed in location\,1. Therefore, we further classified locations\,7 and 1 as two different scenarios.}}
\begin{itemize}
    \item \textbf{Scenario 1 - Rural:} In this scenario, as illustrated in \fref{fig:scenaris_a} and labelled as location\,1, the measurement equipment was placed in a middle of the farm, where neither buildings nor trees were located in a close proximity. As the measurement location was still located close to the city, multiple cells were decodable. This scenario imitates the envisioned use cases of using \acp{UAV} for field inspection.
    \item \textbf{Scenario 2 - Industrial:} In this scenario, as illustrated in \fref{fig:scenaris_a} and labelled as location\,7, the measurement equipment was placed between many tall (up to 30~m), industrial buildings in the industrial part of the city. The measurement spot was surrounded by many elements made of concrete and represents the typical industrial environment, where \acp{UAV} are expected to be used for building inspection or for surveillance services. {In addition, the average height and downtilt of the BS antennas near to the measurement locations in both industrial and rural scenarios were around 33\,m and 6$^\circ$, respectively}.
    \item \textbf{Scenario 3 - Urban:} In this scenario, as illustrated in \fref{fig:scenaris_a} and labelled as locations\,2-6, the measurements were conducted in a couple of locations in a city center where the equipment was surrounded by buildings. Their average height was approximately 20~m. This scenario represents many of the envisioned use cases as for example last mile delivery services. {The average height and downtilt of the \ac{BS} antennas in this area were about 25\,m and 6$^\circ$, respectively.}
    %\item \textbf{Scenario 4 - Urban in a proximity of a fjord:} This scenario is very similar to Scenario~3. The only difference, is the proximity of the fjord. The measurements were taken at the shore of the large water surface. From the other side, the measurement equipment was still surrounded by buildings.
\end{itemize}
\subsubsection*{Measurement procedure}\label{section:procedure}
In each of the measurement locations, the measurement procedure was as follows. The designed testbed was placed inside the steel cage and was lifted by a crane as shown in \fref{fig:photomeas}. The measurements were taken at nine different heights from 0 to 40~m with a 5~m step. At each measurement height, 100\,ms-long measurement snapshot of the \ac{LTE} network was taken five times where the reproducibility of the channel characteristics also serves as a verification of the system functionality.\footnote{\addnewfr{To compromise between the limited hardware performance of data transmission rate from \acp{USRP} to \ac{PXI} and a longer data-length expected, we succeeded to record the real-time downlink signals using a complex sampling rate of 40\,MHz, and the time-length of each snapshot was 100\,ms covering 10 radio frames. More details about the hardware limitations can be found in \cite{8894135}.}} Considering the impact of the steel cage and its attenuation on the received signals, the same procedure was repeated twice. During the first time, the steel cage was mounted with the \ac{UCA} on top of the cage, while during the second time, the \ac{UCA} was located below the cage. Finally, to gather as much diverse data as possible, the same measurements were conducted twice, separately for the two network operators. Therefore, in total 2$\times$2$\times$9$\times$5$\times$7=1260 (corresponding to the number of operators, up/downwards \ac{UCA}, number of measurement heights, number of repeated snapshots, number of measurement locations, respectively) measurement snapshots were recorded.

\section{CIR extraction and Channel \acsp{APDP}\label{sect:cirextraction_and_apdp}}
In this section, we will present the post processing procedure to extract the \acp{CIR} for the measured downlink \ac{UCA} array data. Example \acp{APDP} observed at different heights in different scenarios are also illustrated. The preliminary channel characteristics observed from the \acp{APDP} are discussed. % are discussed by observing the \acp{APDP}.
\subsection{CIR extraction from raw data\label{sect:cirextraction}}
The post processing procedure of extracting \acp{CIR} from the measured downlink raw data is similar to the procedure specified in \cite{8576578,7928479}. The main steps include \textit{i) low-pass filtering} to remove possible out-of-band interferences, \textit{ii) primary synchronization} exploiting the primary synchronization signals, \textit{iii) secondary synchronization} utilizing the secondary synchronization signals and \textit{iv) \ac{CIR} extraction} by applying the inverse Fourier transform to the \acp{CTF} obtained exploiting the cell specific signals. The improvement in this work is that before obtaining time synchronization and detecting physical cells, Bartlett beamforming was firstly conducted by combining the measured data at all the 16 antennas of the \ac{UCA}. Through iteratively adapting the beamforming weights of the \ac{UCA} targeting different/grid directions, the \ac{SINR} of the received signals from cells covered by the current beam was increased due to the mitigation of inter-cell interferences, which led to a better cell detection compared to that of using a single antenna in \cite{8576578}. The step \textit{iv) \ac{CIR} extraction} was then applied to the \ac{UCA} for all the detected cells in individual measurement snapshots. Readers are referred to \cite{8576578} for the detailed description of the steps of raw data processing. For each measurement snapshot lasting 0.1 second,
16$\times$200 \acp{CIR}, i.e. 200 consecutive \acp{CIR} for each of the 16 antennas of the \ac{UCA}, can be extracted from the measured data for each detected cell. To avoid possible problems caused by, e.g. the warming-up stage of the \acp{USRP}, only the latter 16$\times$180 \acp{CIR} were considered for the following investigations. In addition, we denote the array \acp{CIR} extracted for a detected cell in one measurement snapshot as $h(m,t,\tau)$ where $m\in\{1,\cdots,M;M=16\}$, $t$ and $\tau$ indicates the antenna index, time and delay, respectively. The time difference between consecutive \acp{CIR} is 0.5\,ms. The bandwidth of the \ac{CIR} is 18\,MHz with 200 delay sampling points, which corresponds to a maximum observable path delay difference of around 11\,$\mu$s, i.e. 3.33\,km.

\subsection{Example \acp{APDP}}
\begin{figure}
\begin{center}
\subfigure[]{{}\includegraphics[width=0.43\textwidth]{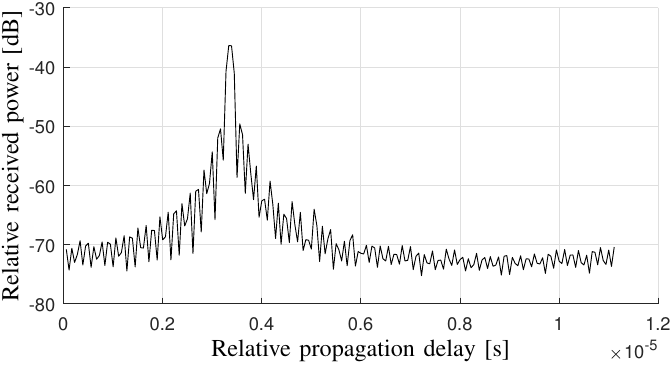}{
\psfrag{Relative propagation delay [s]}[c][c][0.7]{Relative propagation delay [s]}
\psfrag{Relative received power [dB]}[c][c][0.7]{Relative received power [dB]}
}\label{fig:rural_pdp_a}}
\subfigure[]{

	{}\includegraphics[width=0.46\textwidth]{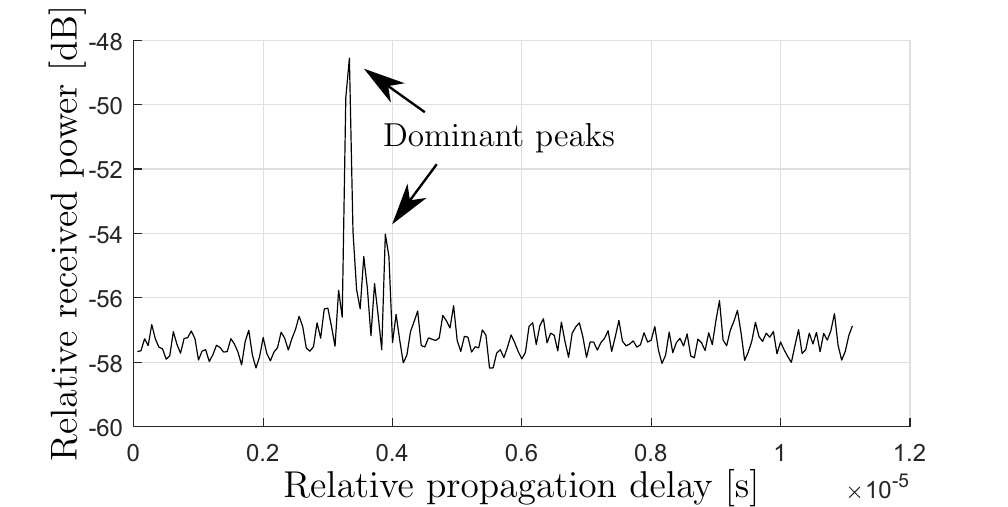}
  
	{\psfrag{Relative propagation delay [s]}[c][c][0.7]{Relative propagation delay [s]}
\psfrag{Relative received power [dB]}[c][c][0.7]{Relative received power [dB]}
}\label{fig:rural_pdp_b}}
\end{center}
\caption{Example \acp{APDP} obtained in the rural scenario in location\,1 as indicated in Fig.\,\ref{fig:photomeas}. (a) Height of 0\,m. (b) Height of 35\,m.  \label{fig:rural_apdps}}
\end{figure}

\begin{figure}
\begin{center}
\subfigure[]{{}\includegraphics[width=0.43\textwidth]{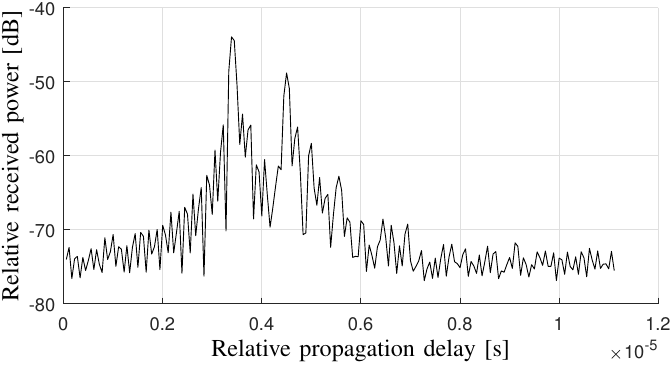}{
\psfrag{Relative propagation delay [s]}[c][c][0.7]{Relative propagation delay [s]}
\psfrag{Relative received power [dB]}[c][c][0.7]{Relative received power [dB]}
}\label{fig:factory_pdp_a}}
\subfigure[]{{}\includegraphics[width=0.43\textwidth]{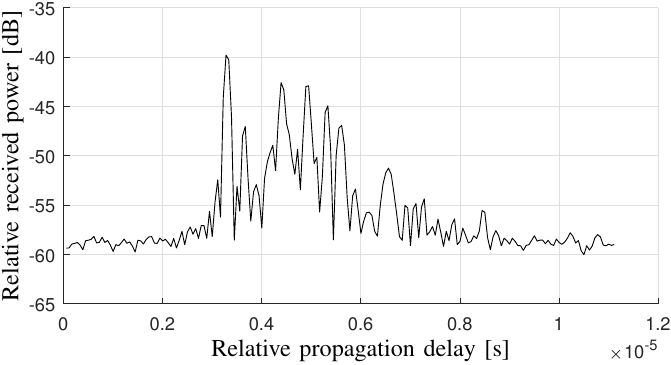}{
\psfrag{Relative propagation delay [s]}[c][c][0.7]{Relative propagation delay [s]}
\psfrag{Relative received power [dB]}[c][c][0.7]{Relative received power [dB]}
}\label{fig:factory_pdp_b}}
\end{center}
\caption{Example \acp{APDP} obtained in the industrial scenario in location\,7 as indicated in Fig.\,\ref{fig:photomeas}. (a) Height of 0\,m. (b) Height of 25\,m.  \label{fig:fac_apdps}}
\end{figure}

\begin{figure}
\begin{center}
\subfigure[]{{}\includegraphics[width=0.43\textwidth]{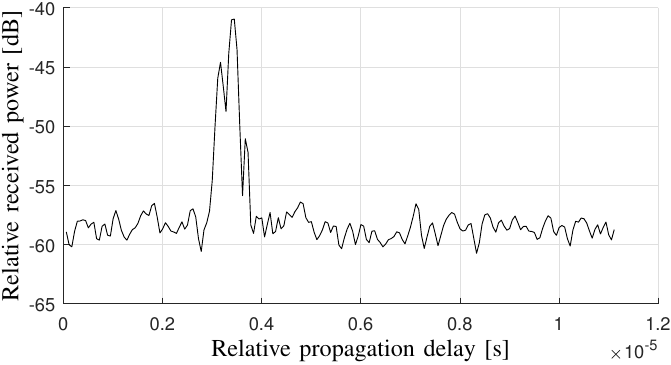}{
\psfrag{Relative propagation delay [s]}[c][c][0.7]{Relative propagation delay [s]}
\psfrag{Relative received power [dB]}[c][c][0.7]{Relative received power [dB]}
}\label{fig:urb_pdp_a}}
\subfigure[]{{}\includegraphics[width=0.43\textwidth]{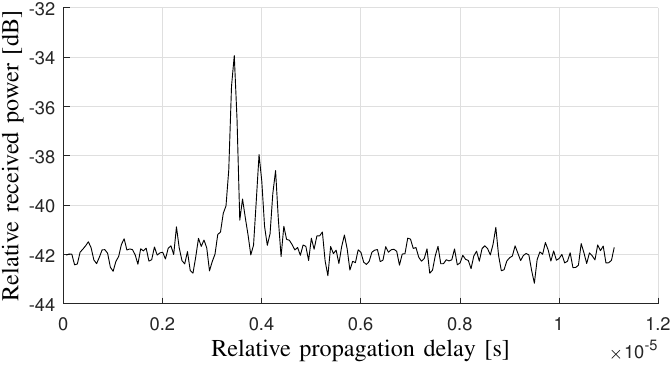}{
\psfrag{Relative propagation delay [s]}[c][c][0.7]{Relative propagation delay [s]}
\psfrag{Relative received power [dB]}[c][c][0.7]{Relative received power [dB]}
}\label{fig:urb_pdp_b}}
\end{center}
\caption{Example \acp{APDP} obtained in the urban scenario in location\,3 as indicated in Fig.\,\ref{fig:photomeas}. (a) Height of 0\,m. (b) Height of 40\,m. \label{fig:urb_apdps}}
\end{figure}

Multiple cells can be detected in one measurement snapshot at a certain height and location. In this section, we show some example \acp{APDP} for different heights at different scenarios. \addnewfr{Note that the illustrated \acp{APDP} may not attributed to the same cell. Thus, in Figs.\,\ref{fig:rural_apdps} to \ref{fig:urb_apdps} the received power of an \ac{APDP} at a higher height is not necessarily higher or lower than that of an \ac{APDP} at a lower height.} The \acp{APDP} are obtained by averaging the \acp{PDP} over time. That is, $s(m,\tau)=\frac{1}{N} \sum_{t_1}^{t_N} |h(m,t,\tau)|^2$ for one set of \ac{UCA} array \acp{CIR} with $N=180$ in our case. For clarity, we only illustrate the \ac{APDP} observed at one antenna of the \ac{UCA} for a detected cell.

\subsubsection{Rural scenario} \fref{fig:rural_apdps} illustrates two \acp{APDP} obtained in the rural area at location\,1 as indicated in \fref{fig:photomeas} at the heights of 0 and 35\,m, respectively. Note that the propagation delays are relative due to the fact that the time synchronization is done offline in the \ac{CIR} extraction. It can be inferred from \fref{fig:rural_pdp_a} that the illustrated channel at the height of 0\,m is dominant with one path (i.e., the \ac{LoS} path), as the \ac{APDP} has a close-to-perfectly symmetric sinc-alike shape\footnote{The reason that we infer it is a \ac{LoS} path is illustrated in \fref{fig:sage_verification}.}. This is reasonable since the rural area is rather open with almost no scatterers such as tall buildings that can lead to significant path components. Moreover, the dynamic range of the \ac{APDP} is high, which is mainly because the cellular \ac{BS} antenna beam is always down tilted towards the ground to better serve the ground users.
\addnewfr{\fref{fig:rural_pdp_b} illustrates another \ac{APDP} at the height of 35\,m with a smaller dynamic range, which is probably because \textit{i)} the \ac{UCA} was out of the main lobe of the \ac{BS}, and \textit{ii)} with less blockage at a higher height, interferences from other cells also increased the noise floor.} However, it is obvious that at least two path components exist in the channel as indicated by the two peaks in \fref{fig:rural_pdp_b}. Moreover, the power difference of the two peaks is not large. This demonstrates the fact the at a higher height, the channel may not always become more \ac{LoS}-alike as assumed in many other works, even in the rural area. Nevertheless, we show this \ac{APDP} as a special case. Most of the channels in the rural scenario became LoS-dominant at higher heights, which we will discuss in details in \sref{sect:channel_models}.

\subsubsection{Industrial scenario} \fref{fig:fac_apdps} illustrates two \acp{APDP} measured in the industrial area at location\,7 as indicated in \fref{fig:photomeas} at the heights of 0 and 25\,m, respectively.
%Although it is near to the rural area, the industrial area with factory buildings existing is not that open compared to the rural area.
As illustrated in \fref{fig:factory_pdp_a}, multiple peaks (or path components) can already be observed in the \ac{APDP} at the height of 0\,m attributed to e.g. the factory buildings. At a higher height, e.g. 25\,m as indicated in \fref{fig:factory_pdp_b}, the richness of the path components of the channel can be even higher. We postulate this is probably due to the reflections from the roofs or round walls of the factory buildings, as indicated in \fref{fig:scenaris_b}, which are visible to the \ac{UAV}-\ac{UE} at a properly higher height.

\subsubsection{Urban scenario} \fref{fig:urb_apdps} illustrates two \acp{APDP} measured in the urban area at location\,3 as indicated in \fref{fig:photomeas} at the heights of 0 and 40\,m, respectively. It can be observed from \fref{fig:urb_pdp_a} that at the height of 0\,m, multiple path components exist in the channel. Moreover, it can be inferred that the \ac{LoS} path has probably been blocked as the path with the highest power is not with the minimum delay. The phenomena are understandable since there are multiple buildings in the dense urban area. \fref{fig:urb_pdp_b} illustrates another example \ac{APDP} at the height of 40\,m, where multiple path components can still be observed. We postulate this is also attributed to the richness of scattering objects in the urban scenario.

To summarize, in this section we have shown some interesting channels measured in different scenarios at different heights. By observing the shape of the \acp{APDP}, some preliminary insights into the temporal characteristics of the channels can be appreciated. The channel at a higher height may not be always \ac{LoS} dominant. Moreover, at a certain height in a specific scenario, e.g. 25\,m at the industrial area, the channel may have higher richness of multipath components, due to certain physical mechanisms such as reflections from roofs and sidewalls. Note that the selected interesting/non-intuitive \acp{APDP} may not represent the general channels. To gain more detailed and general insights into the channels with high resolutions considering also the angular domain, in the sequel we implement a \ac{HRPE} algorithm for all the measured downlink \ac{UCA} array data, and cluster identification will be applied to the estimated channel parameters.

\section{Channel parameter estimation and clustering\label{sect:data_processing}}
In this section, the high resolution estimation of channel parameters (including delays, powers and angles of multipath components) is firstly elaborated.  The practical issues such as calibration and incomplete knowledge of radiation patterns are also considered carefully, which is important for the algorithm to work in practice for the measured data sets. Moreover, the reasonability of the results is not only guaranteed theoretically but also verified empirically. Based on the \ac{HRPE} results obtained, a threshold-based algorithm is applied to group the channels into clusters each consisting of multipath components with similar delays and angles, to facilitate the model establishment.

\subsection{High resolution channel parameter estimation}
\subsubsection{Signal model and physical impairments}

The exploitation of the \ac{UCA} makes it possible to extract the propagation channel parameters in the angular domain from the measured array \acp{CIR}. In this section, we utilize an approach which is derived based on the \ac{SAGE} principle \cite{753729} for this purpose. The \acs{SAGE} algorithm has been widely used for channel parameter estimation due to its low complexity and ability to obtain \ac{HRPE} results \cite{753729}. As a variant of the \ac{EM} algorithm, the \ac{SAGE} algorithm includes E-steps and M-steps to achieve convergence. In the E-step, the conditional expectations of signals contributed by individual path components are calculated based on the prior information obtained in the M-step, where the channel parameters of a path are estimated using successive dimension-wise maximization-likelihood principle, and the estimated parameters are updated immediately for the following E-steps. Readers are referred to \cite{753729,xuesong_tap} for the implementation details.

\begin{figure}
\centering
{}\includegraphics[width=0.43\textwidth]{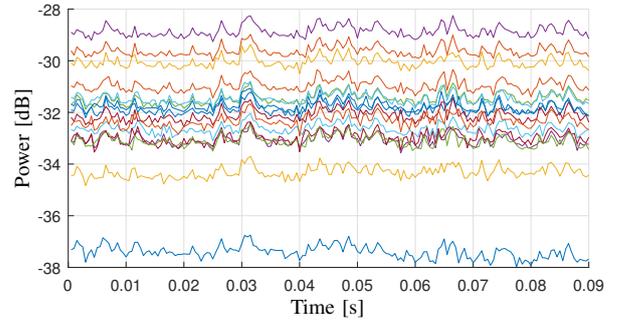}{
\psfrag{Time [s]}[c][c][0.7]{Time [s]}
\psfrag{Power [dB]}[c][c][0.7]{Power [dB]}
}
\caption{The received power at the 16 antennas for the same path component. This example is obtained from a rural data. The different lines represent the received power at the 16 antennas, respectively. \label{fig:r}}
\end{figure}

The underlying signal model assumed for the propagation channel is formulated as
\begin{equation}
\begin{aligned}
 & h_\text{ideal}(t, \tau, \phi, \theta, \nu )  \\
 & = \sum_{\ell=1}^{L(t)} \alpha_\ell(t) \delta(\tau-\tau_\ell(t))     \delta(\phi-\phi_\ell(t)) \delta(\theta-\theta_\ell(t))   \delta(\nu-\nu_\ell(t))
\end{aligned}\label{eq1}
\end{equation} where $L(t)$ is the total number of the propagation paths in the channel snapshot observed at time instant $t$; $\alpha_\ell(t)$, $\tau_\ell(t)$, $\phi_\ell(t)$, $\theta_\ell(t)$ and $\nu_\ell(t)$ represent the complex amplitude, delay, azimuth angle, elevation angle and Doppler frequency of the $\ell$th path component, respectively, and $\delta(\cdot)$ indicates the Dirac delta function. Note that in the measurement campaign, we assume the channel parameters are stationary in each measurement snapshot as the \ac{UCA} was kept almost still when collecting the downlink data. In other words, we assume the path number and channel parameters, e.g. delays and angles, of the propagation channel between a \ac{BS} and the \ac{UCA} do not change during one measurement snapshot. For different measurement snapshots or detected cells  $L$ can be different. The reasons for including Doppler frequencies in the signal model are two folds. One is that small movement of \ac{UCA} caused by, e.g. the wind etc., was inevitable. The other reason is that there may exist frequency shifts in oscillators of \acp{BS} and the \ac{UCA} measurement system, which could cause ``fake'' Doppler frequencies. Therefore, the empirically received \ac{UCA} \acp{CIR} according to \eqref{eq1} with additive white Gaussian noise is formatted as
\begin{equation}
\begin{aligned}
 & h(m,t,\tau) \\ 
 & = \sum_{\ell=1}^{L} \alpha_\ell \mathbf a(\phi_\ell,\theta_\ell) \odot \mathbf{r}_\ell(\phi_\ell,\theta_\ell)  u(\tau-\tau_\ell) e^{j 2\pi \nu_\ell t } + n(m,t,\tau)
\end{aligned}
\label{eq2}.
\end{equation} In \eqref{eq2}, $\mathbf a(\phi_\ell,\theta_\ell)$ is the so-called steering vector, a complex vector of dimension $M\times 1$ that can be readily formatted according to the \ac{UCA} geometry considering the path impinging with azimuth $\phi_\ell$ and elevation $\theta_\ell$. Due to the differences among the antenna radiation patterns and the different blockages for individual antennas caused by e.g. the assemble setup, another vector of dimension
$M\times  1$, $\mathbf{r}_\ell(\phi_\ell,\theta_\ell) \triangleq [r_{\ell,1}(\phi_\ell,\theta_\ell), \cdots, r_{\ell,M}(\phi_\ell,\theta_\ell)]^{T}$ has to be included to compensate the received power differences among antennas. $T$ and $\odot$ indicate the matrix transpose operation and Hadamard product (element-wise multiplication), respectively. Moreover, $u(\tau)$ is the shape function introduced by the measurement system. Ideally, $u(\tau)$
is a sinc function if there is only the effect of limited bandwidth (18\,MHz in our case). However, the measurement system usually has other system imperfections, which compositely makes $u(\tau)$ very different from an ideal sinc function. As for $\mathbf r_\ell$, \fref{fig:r} illustrates the received powers of the dominant path\footnote{The power of the dominant peak of each \ac{CIR}, e.g. as indicated in \fref{fig:rural_pdp_a}, is collected.} at the 16 antennas, obtained from a rural (close to \ac{LoS}) $h(m,t,\tau)$ as in \eqref{eq2}. \addnewfr{Ideally, the powers of the LoS path received at different antennas should be the same or with little difference. However, it can be observed from  \fref{fig:r} that the received powers at different antennas can be very different, which we postulate is mainly due to the radiation pattern difference and blockage difference across the \ac{UCA} elements. This imperfection problem has to be considered when performing the parameter estimation. }
%It can be observed that the power difference among antennas can be significant, which we postulate is mainly due to the radiation pattern difference and blockage difference across the \ac{UCA} elements. 
One may think that this can be calibrated by measuring the patterns in a anechoic chamber. However, the difficulty is to make sure the \ac{UCA} has the same status in the chamber to that in the measurement campaign. The rotation must refer to the phase center of the \ac{UCA} (which is physically unknown), and the other 15 antennas should be kept active when one element is measured. The practical difficulties make it almost impossible to obtain good calibration results of
$\mathbf r_\ell$ that can improve the estimation. The model mismatch will lead to failures both in the E-steps and M-steps, and results show that the original \ac{SAGE} algorithm does not work as many more spurious (ghost) path-components has been estimated if $\mathbf r$ is omitted as $\mathbf 1$-vector. This mean that we must carefully consider the physical impairments, i.e.
$u(\tau)$ and $\mathbf r_\ell$, in the parameter estimation to obtain realistic results.

\subsubsection{Calibrate $u(\tau)$} 
\begin{figure}
\centering
{}\includegraphics[width=0.43\textwidth]{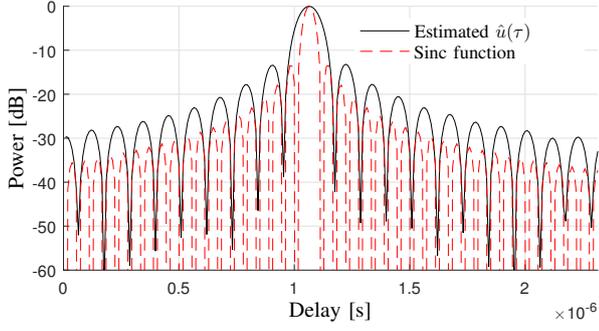}{
\psfrag{Delay [s]}[c][c][0.7]{Delay [s]}
\psfrag{Power [dB]}[c][c][0.7]{Power [dB]}
\psfrag{Sinc function}[l][l][0.6]{Sinc function}
\psfrag{Estimated shape function}[l][l][0.6]{Estimated $\hat u(\tau)$}
}
\caption{The shape of the estimated $u(\tau)$ of the \ac{USRP} and the shape of standard bandwidth-limited sinc function.\label{fig:u_tau}}
\end{figure}

The following operations were conducted in order to obtain $u(\tau)$ as close as possible to that in the field measurements. The same version of the standard \ac{LTE} \ac{FDD} signals transmitted by the \acp{BS} in the measurement campaign were generated. The generated signals were then fed to and transmitted by {\red a} \ac{USRP} in the laboratory. The type of the \ac{USRP} was the same to that used for the \ac{UCA}, and the center frequencies and bandwidth were set identical to that of the commercial networks, respectively. Moreover, the Tx \ac{RF} port of this \ac{USRP} was directly connected to the Rx \ac{RF} port of the primary \ac{USRP} (which was connected to the first antenna of the \ac{UCA} in the measurements). The cable used in the field measurements was still used for the direct connection with {a} necessary attenuator added. The transmitted ``downlink \ac{LTE} signals'' were then collected by the primary \ac{USRP} using the exactly same configuration to that for sounding the live \ac{LTE} signals. Further, we exploited the same procedure as elaborated in \sref{sect:cirextraction} to extract the \acp{CIR}
$h_{\text{d}}(t,\tau)$ from the received data. To remove the effect of thermal noise and oscillator inaccuracy in the \ac{USRP}, we firstly estimated the Doppler frequency caused by the oscillator inaccuracy as\footnote{\addnewfr{Note that the two \acp{USRP} were both static when conducting calibration, thus the phase rotation (if any) between neighboring \acp{CIR}  was only introduced by the center frequency offset between the two \acp{USRP}.}}
\begin{equation}
\begin{aligned}
\hat{\nu}_{\text{d}}  =  \arg \max_{\nu} \left| \sum_{t=t_1}^{t_{N_\text{d}}} e^{-j2\pi \nu t} h_{\text{d}}(t,\hat\tau_{\text{d}}) \right|^2,
\end{aligned}
\end{equation}
%with
\begin{equation}
\begin{aligned}
\hat{\tau}_{\text{d}}  =  \arg \max_{\tau} \sum_{t=t_1}^{t_{N_\text{d}}} \left| h_{\text{d}}(t,\tau)\right|^2
\end{aligned}
\end{equation} where $N_\text{d}$ is the number of \acp{CIR} utilized. Then $u(\tau)$ is estimated by coherently combining the \acp{CIR} as
\begin{equation}
\begin{aligned}
\hat u(\tau)  =  \sum_{t=t_1}^{t_{N_\text{d}}} h_{\text{d}}(t,\tau) e^{-j2\pi \nu t}
\end{aligned}.
\end{equation} Normalization of power and phase and shifting delay to 0 were also applied afterwards. \fref{fig:u_tau} illustrates the estimated $\hat u(\tau)$ using $N_\text{d}=400$ \acp{CIR}. The ``fake'' Doppler frequency was estimated as around -12\,Hz. A standard 18\,MHz bandwidth-limited sinc function is also plotted. Their delays are shifted from 0 for a better presentation in \fref{fig:u_tau}. It can be observed that the system effects of the \ac{USRP} utilized in the measurements is indeed non-negligible. The obtained close-to-noise-free $\hat u(\tau)$ makes it possible to remove/calibrate the system effects of $u(\tau)$.

\subsubsection{Deal with unknown $\mathbf r_\ell$}: A novel strategy is proposed to deal with the unknown $\mathbf r_\ell$. The main idea is to firstly estimate only amplitudes, delays and Doppler frequencies of path components at individual antennas, and angular parameters are then estimated exploiting the complex amplitudes of 16 antennas. Specifically, in the E-step, the expectation of the hidden signals contributed by the $\ell$th path and received by the $m$th antenna is calculated as
\begin{equation}
\begin{aligned}
\hat h_{m,\ell}(t,\tau)  = h(m,t,\tau) - \sum_{q\in \mathcal{L}, q \neq \ell} \alpha_{m,q}^\prime u(\tau-\tau_q^\prime) e^{j2\pi\nu_q^\prime t}
\end{aligned}
\end{equation} where $\alpha_{m,q}^\prime$,  $\tau_q^\prime$ and $\nu_q^\prime$ are the estimated complex amplitude, delay and Doppler frequency respectively in the previous M-step for the $q$th path received at the $m$th antenna, and $\mathcal{L} = \{1,\cdots,L\}$ is the set of all path indices. In the next M-step, the maximum-likelihood estimations are formatted as
\begin{equation}
	\scalemath{0.81}{
\begin{aligned}
(\tau_\ell^\prime, \nu_\ell^\prime) =  \arg \max_{\tau_0, \nu}  \left\{\sum_{m=1}^{M}   \left| \sum_{t_1}^{t_N} e^{-j2\pi\nu t}  \int u^{*} (\tau-\tau_0)  h_{m,\ell}(t,\tau) d\tau \right|^2 \right\}
\end{aligned}}
\end{equation}
\begin{equation}
\begin{aligned}
 \alpha_{m,\ell}^\prime  = \frac{1}{N\int |u(\tau)|^2 d\tau} \sum_{t_1}^{t_N} e^{-j2\pi\nu_\ell^\prime t} \int u^{*} (\tau-\tau_\ell^\prime)  h_{m,\ell}(t,\tau) d\tau
\end{aligned}
\end{equation} where $^{*}$ indicates the conjugate of the argument. With convergence, the parameters $\boldsymbol{\hat \alpha}_\ell \triangleq [\hat\alpha_{m,1}, \cdots,\hat\alpha_{m,\ell}]^{T}$, $\hat\tau_\ell$ and $\hat\nu_\ell$ are obtained. The azimuths, elevations and amplitudes of individual paths are estimated as
\begin{equation}
\begin{aligned}
(\hat \phi_\ell, \hat\theta_\ell) = \frac{1}{M} \arg \max_{\phi,\theta} \boldsymbol{\hat \alpha}_{\ell}^{T} \mathbf a^{*}(\phi, \theta)
\end{aligned},
\end{equation}
\begin{equation}
\begin{aligned}
\hat\alpha_\ell = \frac{1}{M} \boldsymbol{\hat \alpha}_{\ell}^{T} \mathbf a^{*}(\hat\phi_\ell, \hat\theta_\ell)
\end{aligned}.
\end{equation}
The final parameter set obtained for one measurement snapshot is denoted as $\hat \bsTheta = [\hat\alpha_\ell, \hat\tau_\ell, \hat\phi_\ell, \hat\theta_\ell; \ell \in \mathcal{L} ]$. Moreover, it is essential to determine an appropriate $L$ (that can change for different snapshots or cells) to avoid over-estimation. The \ac{AIC} principle \cite{1100705,8412215} is exploited where the $L$ that leads to the minimum \ac{AIC} is used, which is equivalently to find the number of estimated paths with \ac{SNR} above a certain threshold \cite{8412215}. Therefore, in practice an adequately large number of paths are firstly estimated. $L$ is chosen as the number of paths with powers above the noise floor, and the estimation is reperformed with the determined $L$ to obtain final estimation results.

\begin{figure}
\begin{center}
\subfigure[]{{}{}\includegraphics[width=0.42\textwidth]{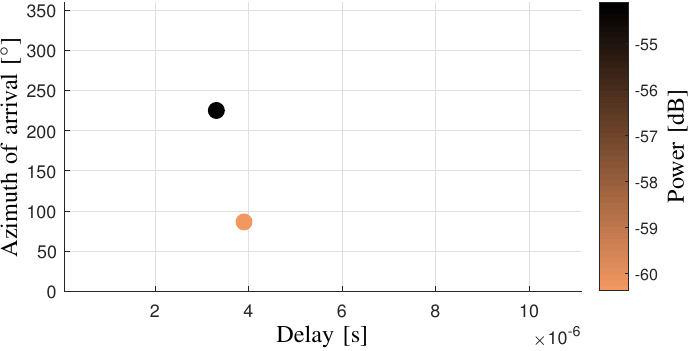}{
\psfrag{Delay [s]}[c][c][0.7]{Delay [s]}
\psfrag{Azimuth [deg]}[c][c][0.7]{Azimuth of arrival [$^\circ$]}
\psfrag{Power [dB]}[c][c][0.7]{Power [dB]}
}\label{fig:sage_a}}
\subfigure[]{{}\includegraphics[width=0.42\textwidth]{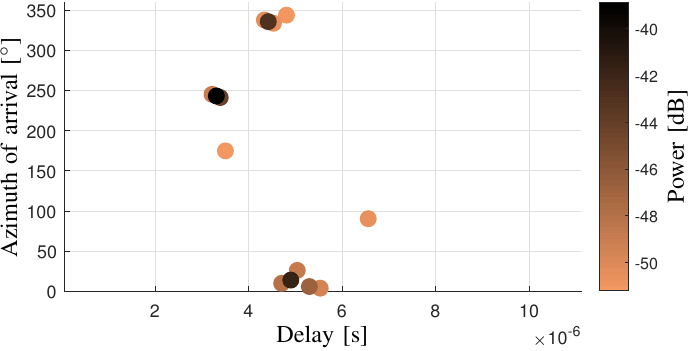}{
\psfrag{Delay [s]}[c][c][0.7]{Delay [s]}
\psfrag{Azimuth [deg]}[c][c][0.7]{Azimuth of arrival [$^\circ$]}
\psfrag{Power [dB]}[c][c][0.7]{Power [dB]}
}\label{fig:sage_b}}
\end{center}
\caption{Example \ac{HRPE} results for the measured channels. (a) The rural channel as indicated in \fref{fig:rural_pdp_b}. (b) The industrial channel as indicated in \fref{fig:factory_pdp_b}. \label{fig:sage_estimates}}
\end{figure}

\begin{figure}
\centering
\includegraphics[width=0.43\textwidth]{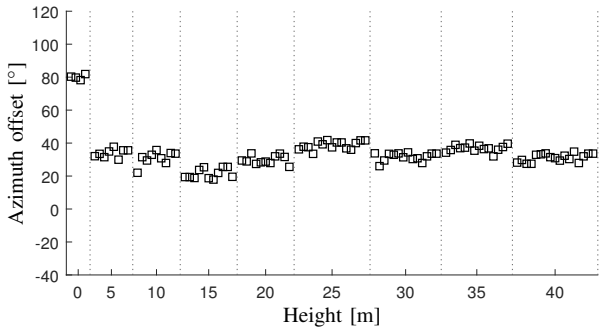}{
\psfrag{Delay [s]}[c][c][0.7]{Delay [s]}
\psfrag{Angle diff}[c][c][0.7]{Azimuth offset [$^\circ$]}
\psfrag{Height [m]}[c][c][0.7]{Height [m]}
\psfrag{up}[l][l][0.6]{Up-{UCA}}
\psfrag{down}[l][l][0.6]{Down-{UCA}}
}
\caption{Offsets between the dominant azimuths estimated and the corresponding geographically \ac{LoS} azimuths for individual cells detected in the rural scenario. The vertical dotted-lines separate the cells detected at different heights.\label{fig:sage_verification}}
\end{figure}

\fref{fig:sage_estimates} illustrates two example \ac{HRPE} estimation results for the \ac{UCA} \acp{CIR} snapshots in \fref{fig:rural_pdp_b} and \fref{fig:factory_pdp_b}, respectively. It can be observed that the number of paths estimated in the rural example is much lower than that of the industrial example. Moreover, in the rural example, the two paths have no surrounding paths in their near regions in terms of delay and angle. As a comparison, groups of paths with similar delays and angles can be found in \fref{fig:factory_pdp_b} in the industrial example, although the well-separated single paths also exist. This is consistent with the intuition that richer number of scatters can result in a richer amount of paths, and sometimes a surface or open area can lead to a single path considering the finite intrinsic resolution ability achieved by the measurement system.
To further empirically check the estimation results, as illustrated in \fref{fig:sage_verification}, we compared the azimuth angles of the dominant paths estimated for individual cells detected at different heights and the corresponding geographical \ac{LoS} azimuth angles (with north direction as 0$^\circ$) calculated according to the physical locations of \acp{BS} and the \ac{UCA} in the rural scenario.
%We assume that the \ac{AoA} of the dominant path of the serving cell should be constant among all snapshots received at the same height regardless of the antenna orientation.
It can be observed from \fref{fig:sage_verification} that the offsets are rather stable inside each height. The change between different heights is due to the practically inevitable rotation of the \ac{UCA} when it was lifted up to another height. Therefore, we do consider the \ac{HRPE} results obtained in our work are realistic in both the theoretical and empirical senses. %This is the sanity check of our assumptions regarding channel stationarity as well as the \ac{SAGE} implementation.%  are around 180$^\circ$. This is consistent with what we tried in the measurements, i.e., to keep the orientations of the \ac{UCA} the same for all heights and do not rotate the \ac{UCA} when turning it upside down. Note that although there is a ``jumping'' from 0\,m to 5\,m, all the offsets at 0\,m are still rather stable. Therefore, we do consider the \ac{HRPE} results are feasible in our work theoretically and empirically.

\subsection{Cluster identification}
The concept of channel clustering has been widely used to compromise between the model complexity and accuracy of spatial channels. A cluster is considered as a group of multipath components with similar delays and/or angles. %As an example, the different groups of paths in \fref{fig:sage_b}.
It is essential to utilize an automatic algorithm to identify the clusters of a large amount of spatial channels available. Different approaches, e.g. \ac{KPM} \cite{4109300,4469589}, image-processing-based methods \cite{8509190,8240983} and several others \cite{book:275224,8103059} can be found in the literature. In this work, we utilize the \ac{MCD}-threshold based principle \cite{6691924}.
%Different approaches can be found in the literature. \Ac{KPM} algorithm \cite{4109300,4469589}, as one of the earliest and representative clustering algorithm, includes the powers in the calculation of \ac{MCD}, which was improved based on the KMeans algorithm \cite{book:275224}. However, the initializations of cluster centroids and cluster number to be identified affect the performance of \ac{KPM} significantly. In \cite{8103059}, by assuming the paths of a cluster obey multivariate Gaussian distribution, a \ac{GMM} was proposed to group paths. Similarly, in \cite{8013075}, a Kernel-Power-Density algorithm has been proposed where Gaussian and Laplacian kernel-densities were assumed for delays and angles, respectively. In \cite{8509190}, an image-processing-based method was proposed to deal with (non-high-resolution) \ac{PAS} images for cluster identification, and \ac{PDP} images were exploited in \cite{8240983} to identify evolving clusters in the delay domain. Moreover, a \ac{MCD}-threshold based algorithm was proposed in \cite{6691924}. We also utilize the threshold-based principle in our work.
The reasons are as follows: \textit{i)} Based on the observation of the
\ac{HRPE} results obtained for the measured channels, it is found that different path-groups are quite well separated with paths confined in a certain delay-angle region as exemplified in \fref{fig:sage_b}. \textit{ii)} In the threshold-based algorithm, no initializations or prior assumptions of cluster centroids, cluster number, paths distributions, etc. are required. The optimum threshold is physically linked to the cluster size or distribution. The investigations in \cite{7331737,8103059} have also shown the performance enhancement of the threshold-based algorithm compared to \ac{GMM} \cite{8103059}, \ac{KPM} \cite{4109300} and KMeans \cite{book:275224}.

The \ac{MCD} has been exploited to replace the Euclidean distances in cluster identification due to its significant performance improvement \cite{1577605}, which was firstly introduced and discussed in \cite{MCD} to quantify the separations of multiple paths in multiple parameter domains. The \acp{MCD} in delay and angular domains are calculated differently. Specifically, the angular \ac{MCD} between the $i$th and $j$th path components is calculated as
\begin{equation}
\begin{aligned}
 \text{MCD}_{\text{Tx/Rx},ij} = \frac{1}{2}\Biggl | \begin{pmatrix}
\sin\theta_i\cos\phi_i  \\
\sin\theta_i\sin\phi_i\\
\cos\theta_i
\end{pmatrix} - \begin{pmatrix}
\sin\theta_j\cos\phi_j  \\
\sin\theta_j\sin\phi_j\\
\cos\theta_j
\end{pmatrix} \Biggl |
 \end{aligned}
 \label{eq:mcd1}
\end{equation} for Tx and Rx sides, separately. The delay \ac{MCD} is calculated as
\begin{equation}
\begin{aligned}
 \text{MCD}_{\tau,ij} = \frac{|\tau_i - \tau_j|}{\tau_\zeta}
\end{aligned}
\label{eq:mcd2}
\end{equation} with $\tau_\zeta$ as a scaling factor for delay. Note that the calculation in \eqref{eq:mcd2} is different from the original/widely-used definition of delay \ac{MCD}, e.g. in \cite{4109300,7331737}. In the original definition, the scaling factor is subject to the maximum difference and standard deviation of path delays in a channel snapshot. This means that the scaling factor can change arbitrarily, e.g., if a cluster with a long delay exists or all the clusters (well separated in angular domain) have similar delays. However, in the same scenario clusters probably have similar physical sizes, thus it may lead to unrealistic identification results. In \eqref{eq:mcd2},  $\tau_\zeta$ is selected to make the angular \ac{MCD} and delay \ac{MCD} of the maximum angle and delay separateness inside clusters of the same scenario be the similar/same magnitudes, which is consistent with the threshold principle.\footnote{This operation is linked to the underlying physical mechanisms. For example, we observe that in most cases of a scenario the maximum azimuth difference and delay difference inside clusters are around 30$^\circ$ and 2\,$\mu$s. Then $\tau_\zeta$ is set to achieve equality between the angular \ac{MCD} of \eqref{eq:mcd1} with angle separateness of
30$^\circ$ and the delay \ac{MCD} of \eqref{eq:mcd2} with delay separateness of 2\,$\mu$s.} The overall \ac{MCD} is then calculated as
\begin{equation}
\begin{aligned}
\text{MCD}_{ij} = \sqrt{ \text{MCD}_{\text{Tx},ij}^2 + \text{MCD}_{\text{Rx},ij}^2 + \text{MCD}^2_{\tau,ij}}
\end{aligned}.
\label{eq:mcd3}
\end{equation} \addnewfr{It is worth noting that due to the low elevation-sensitivity of \ac{UCA} and the fact that the horizontal distances between the UCA and detected cells are gernerall large causing similar and small elevation angles of paths, we omit $\theta$ in the calculation of \eqref{eq:mcd1}.} Moreover, since the angular information at the \ac{BS} side is unknown, we also omit
$\text{MCD}_{\text{Tx},ij}$ in \eqref{eq:mcd3}. In addition, the centroid $\mu_c \triangleq [\mu_\tau, \mu_\phi]$ of a cluster is calculated as
\begin{equation}
\begin{aligned}
\mu_\tau =  \frac{\sum_{\ell\in\mathcal{L}_c}|\alpha_\ell|^2   \tau_\ell} {\sum_{\ell\in\mathcal{L}_c}|\alpha_\ell|^2},\quad \mu_\phi = \text{ang}\{ {\sum_{\ell\in\mathcal{L}_c}|\alpha_\ell|^2   e^{j\phi_\ell}} \}
\end{aligned}
\label{eq:cluster_center}
\end{equation} where $\mathcal{L}_c$ is the set containing the path indices belong to the concerned cluster, and $\text{ang}\{\cdot\}$ is the phase angle of the argument.

\begin{figure}
\centering
{}\includegraphics[width=0.43\textwidth]{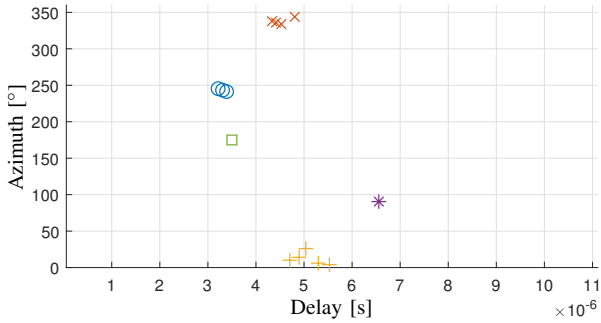}{
\psfrag{Delay [s]}[c][c][0.7]{Delay [s]}
\psfrag{Azimuth [deg]}[c][c][0.7]{Azimuth [$^\circ$]}
}
\caption{Identified clusters for the channel as illustrated in \fref{fig:sage_b}. \label{fig:clusters_identified}}
\end{figure}

\begin{figure*}
	\centering
	{
	{
			{}\includegraphics[width=0.75\textwidth]{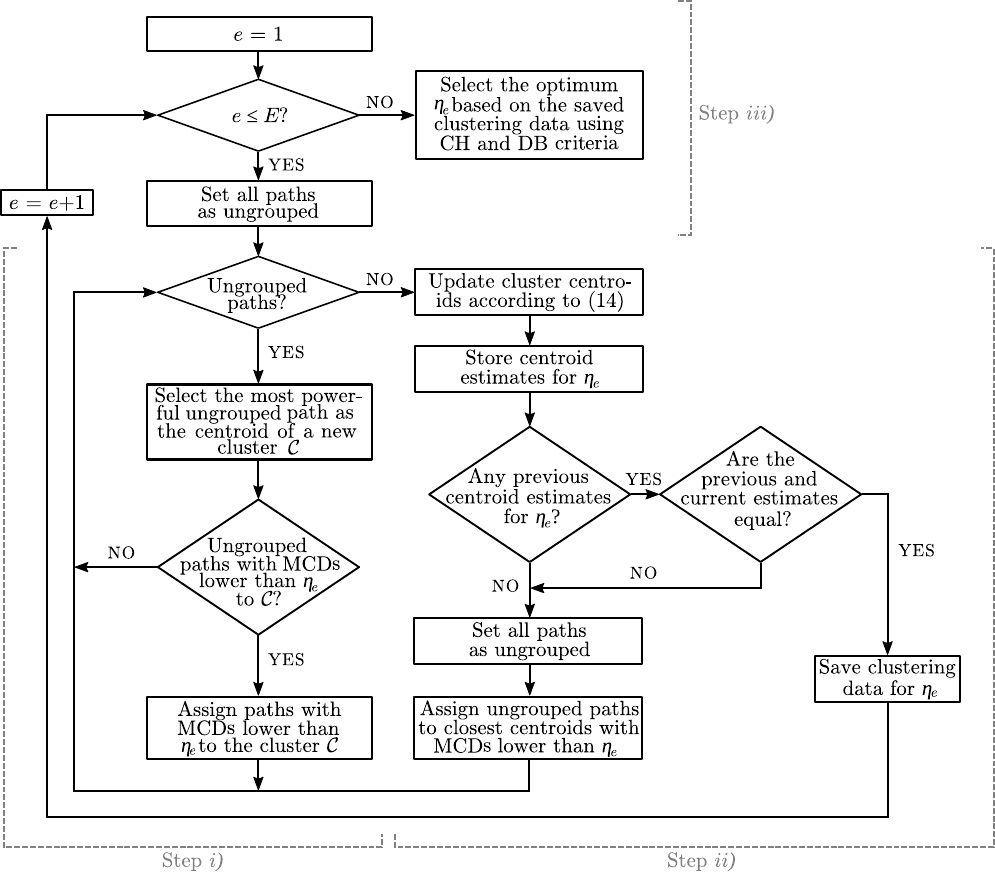}
	}}
	\caption{Cluster identification procedure applied. \label{al1}}
\end{figure*}

The \addnewfr{block diagram} in \addnewfr{Fig.}\,\ref{al1} illustrates the threshold-based cluster identification for the estimated $\hat \bsTheta$, which is similar to the procedure as proposed in \cite{9115069}. Briefly\addnewfr{\footnote{\addnewfr{Note that for convenience the steps \textit{i)}, \textit{ii)} and \textit{iii)} are indicated in the block diagram on Fig.\,\ref{al1}.}}}, \textit{i)} the path component with the highest power is chosen as the first cluster centroid, and the path components with \acp{MCD} less or equal to the pre-defined threshold \addnewfr{$\eta_{e}$} are considered as belonging to this cluster. The same procedure continues for the residual paths until all the paths have been concerned. \textit{ii)} Then the centroids of the grouped clusters are updated using \eqref{eq:cluster_center}, and the threshold-based path-assignment is re-performed referring to the updated cluster centroids. Note that \textit{i)} is also required if there exist paths belonging to none of the updated centroids. The identified clusters are obtained for this \addnewfr{$\eta_{e}$} after several iterations achieving convergence. \textit{iii)} It is essential to determine
\addnewfr{$\eta_{e}$} for the cluster identification. We exploit the same approach as proposed in \cite{4109300} to find the optimum \addnewfr{$\eta_{e}$}, i.e. the optimum cluster identification. That is, \textit{i)} and \textit{ii)} are performed for different candidate \addnewfr{$\eta_{e}$, with $e=1,2,\dots,E$}. The optimum \addnewfr{$\eta_e$} is chosen as the one that leads to the optimum cluster identification, by evaluating the inter-cluster separateness and intra-cluster compactness using the \ac{CH} index and \ac{DB} criterion.  \fref{fig:clusters_identified} illustrates the identified clusters for the example channel as illustrated in
\fref{fig:sage_b}. It can be observed that multiple clusters are well identified. Note that we still term the ``cluster'' with only one path as a ``cluster'' for convenience, although clusters are usually considered with multiple path components. Based on the cluster identification results, spatial channel models will be elaborated in the sequel.

%\begin{algorithm}%\captionsetup{labelfont={sc,bf}, labelsep=newline}
%\setstretch{1}
%{\textbf {Input}: \ac{HRPE} channel parameters $\hat \bsTheta$\\}
%{\textbf {Output}: The optimum cluster identification.}

%\begin{algorithmic}[1]
%  \FOR {$\eta_\text{th} \in [\eta_\text{min},  \eta_\text{max}]$}
%        \STATE Initialize clusters as specified in \textit{i)}.
%        \STATE \textbf{Repeat:}
%        \STATE Update cluster centroids and regroup paths.
%        \STATE \textbf{Until} the cluster centroids remain unchanged.
%          \ENDFOR
%  \STATE Find the optimum $\eta_\text{th}$, i.e. optimum identified clusters, according to the \ac{CH} index and \ac{DB} criterion \cite{4109300}.
% \caption{The cluster identification procedure applied. \label{al1}}
%\end{algorithmic}
%\end{algorithm}

\section{Spatial Channel Characteristics\label{sect:channel_models}}
Based on the cluster identification results, the spatial channel characteristics extracted in different scenarios are illustrated in this section. The composite parameters, e.g. the composite delay spread and azimuth spread, as well as cluster-level parameters, e.g. the number of clusters, intra-cluster delay spread, intra-cluster azimuth spread, cluster power ratio, etc. are thoroughly investigated. For each parameter, a ``boxplot'' figure, e.g. \fref{fig:com_delay_spread}, is used to show the parameter variations at different heights in the three scenarios.\footnote{\addnewfr{The bottom and top of each box are the 25th and 75th percentiles of the sample, respectively. The red line in the middle of each box is the sample median. The whiskers are lines extending above and below each box and showing the minimum and maximum of the whole data set.  Observations beyond the whisker length are marked as outliers (red dots). Readers are referred to the documentation of MATLAB function ``boxplot'' for more detailed information.}}
%For each parameter, a ``boxplot'' figure, e.g. \fref{fig:com_delay_spread}, is used to show the parameter variations at different heights in the three scenarios, where the 25-percentile, median and 75-percentile are indicated by the box. 
The parameter set is consistent with the standard spatial channel modelling methodology as specified in \cite{3GPP38901}. For example, with composite spreads, cluster number, cluster power ratio and cluster offsets known, locations and relative powers of individual clusters can be generated. According to the intra-cluster spreads, the path distributions inside clusters can be further reproduced. Reader are referred to \cite{3GPP38901} for detailed procedures of channel reproduction stochastically.

%The concerned parameters include composite ones, e.g. the composite delay spread and azimuth spread, as well as cluster-level parameter, e.g. the number of cluster, intra-cluster delay spread, intra-cluster azimuth spread, cluster power ratio, etc. For each parameter, a ``boxplot'' figure, e.g. \fref{fig:com_delay_spread}, is used to show the parameter variations at different heights in the three scenarios, where the 25-percentile, median and 75-percentile are indicated by the box. %Moreover, a table summarizing all the channel characteristics with more details is presented in the end.

\subsection{Composite delay spread $\sigma_\tau$ and azimuth spread $\sigma_\phi$} As commonly characterized \cite{7928479}, the \ac{RMS} delay spread of a channel can be calculated according to the \ac{HRPE} $\hat \bsTheta$ as
\begin{equation}
\begin{aligned}
\sigma_{\tau}=\sqrt{\overline{{\tau}^2}-\overline{\tau}^2}
\end{aligned}\label{eq:delayspread}
\end{equation}
with
\begin{equation}
\begin{aligned}
\overline{{\tau}^2}=\frac{\sum_{\ell\in \mathcal{L} } \left|\alpha_\ell\right|^2 \cdot\tau_\ell^2}{\sum_{\ell\in \mathcal{L} }\left|\alpha_\ell\right|^2}, \quad \overline{\tau}=\frac{\sum_{\ell\in \{\mathcal{L} \}} \left|\alpha_\ell\right|^2 \cdot\tau_\ell}{\sum_{\ell\in \{\mathcal{L} \}} \left|\alpha_\ell\right|^2}
\end{aligned}.
\end{equation} The \ac{RMS} azimuth spread is calculated differently as specified in \cite{3GPP38901}
\begin{equation}
\begin{aligned}
\sigma_{\phi}=\sqrt{-2\mathrm{log}\left(\left| \frac{\sum_{\ell\in \mathcal{L} } \mathrm{exp}(j\phi_\ell)\cdot\left| \alpha_\ell\right|^2}{\sum_{\ell\in \mathcal{L} } \left| \alpha_\ell\right|^2 }\right|\right)}
\end{aligned}\label{eq:azimuthspread}.
\end{equation}

\begin{figure}
\begin{center}
\subfigure[]{{}\includegraphics[width=0.42\textwidth]{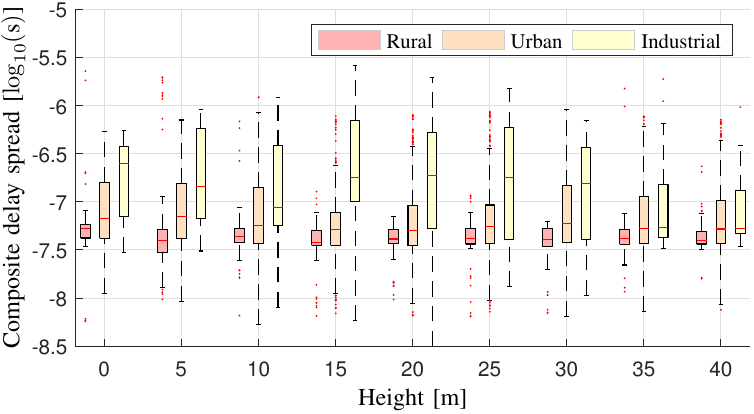}{
\psfrag{Comdelayspread [log10(s)]}[c][c][0.7]{Composite delay spread [$\log_{10}(\text{s})$]}
\psfrag{Height [m]}[c][c][0.7]{Height [m]}
\psfrag{Urban}[l][l][0.6]{Urban}
\psfrag{Rural}[l][l][0.6]{Rural}
\psfrag{Industrialo}[l][l][0.6]{Industrial}
}\label{fig:com_delay_spread}}
\subfigure[]{{}\includegraphics[width=0.42\textwidth]{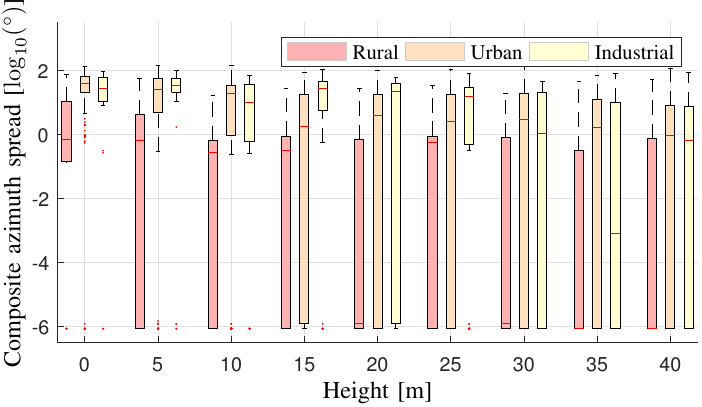}{
\psfrag{Comazimuthspread [log10(deg)]}[c][c][0.7]{Composite azimuth spread [$\log_{10}(^\circ)$]}
\psfrag{Height [m]}[c][c][0.7]{Height [m]}
\psfrag{Urban}[l][l][0.6]{Urban}
\psfrag{Rural}[l][l][0.6]{Rural}
\psfrag{Industrialo}[l][l][0.6]{Industrial}
}\label{fig:com_azimuth_spread}}
\end{center}
\caption{The composite delay spreads and composite azimuth spreads observed in the three scenarios. (a) Composite delay spreads $\sigma_\tau$. (b) Composite azimuth spreads $\sigma_\phi$. \label{fig:comspreads}}
\end{figure}

\begin{figure}
\centering
{
	{
{}\includegraphics[width=0.45\textwidth]{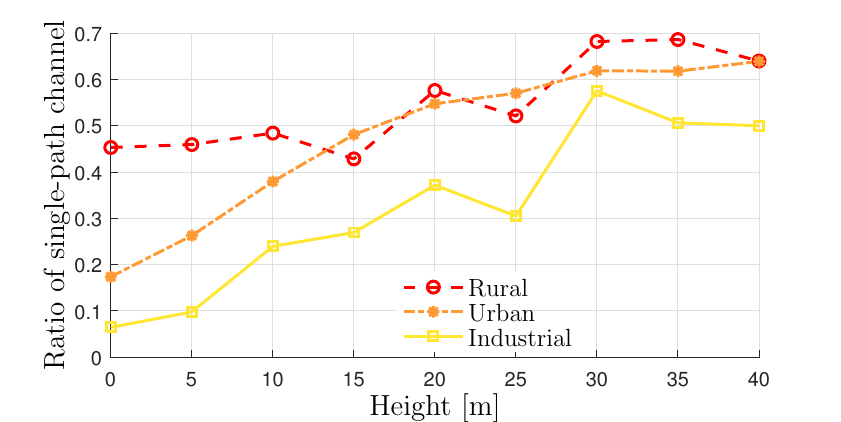}
}}{
\psfrag{Height [m]}[c][c][0.7]{Height [m]}
\psfrag{ratio of channels with one path}[c][c][0.7]{Ratio of single-path channel}
\psfrag{data1}[l][l][0.6]{Rural}
\psfrag{data2}[l][l][0.6]{Urban}
\psfrag{data3}[l][l][0.6]{Industrial}
}
\caption{Ratios of single-path channels at different heights in the three scenarios. \label{fig:ratio_one_path_com}}
\end{figure}

 \fref{fig:comspreads} illustrates the observed composite delay and azimuth spreads, both indicated in logarithm scales, for the three scenarios. Note that a channel may be estimated with only one path whose spreads are 0, therefore such \acp{CIR} are excluded when plotting \fref{fig:comspreads}. The ratios of the single-path channels, defined as the ratio of the number of the channels with only one path to the number of all the channels, at different heights in the three scenarios are illustrated in \fref{fig:ratio_one_path_com}. It can be observed that generally the  composite spreads in the rural scenario are the smallest, and the ratios of single-path channels are the largest (with a ratio of 45\% already at 0\,m). This is consistent with the fact that the rural scenario is quite open. As a contrast, the industrial scenario basically has the largest spreads and smallest ratios of single-path channels, due to the many concrete elements existing. Moreover, it can be observed that the composite spreads in one scenario have the trends to be smaller at a higher height, although they can be larger at some higher heights, e.g. 25\,m in the industrial scenario and 30\,m in the urban scenario. Meanwhile, the ratio of single path channel becomes larger with increasing heights. The phenomena are reasonable since a higher height can in principle result in a less complicated channel, whereas the signals from the building roofs or sidewalls may cause larger spreads at a proper height in the urban and industrial scenarios.

\subsection{Cluster power ratio $K$ and number of clusters $C$}

\begin{figure}%[!tbp]
  \centering
 
		{}\includegraphics[width=0.43\textwidth]{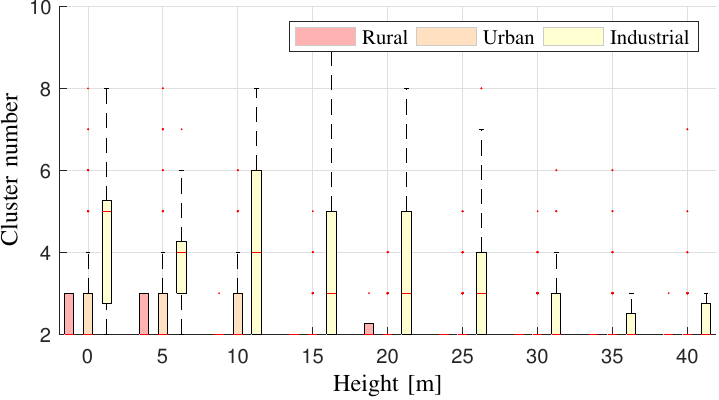}{
		\psfrag{cluster number}[c][c][0.7]{Cluster number}
		\psfrag{Height [m]}[c][c][0.7]{Height [m]}
		\psfrag{Urban}[l][l][0.6]{Urban}
		\psfrag{Rural}[l][l][0.6]{Rural}
		\psfrag{Industrialo}[l][l][0.6]{Industrial}
		}
		\caption{The numbers of clusters $C$ for multiple-clusters channels at different heights in the three scenarios. \label{fig:cluster_number}}
\end{figure}

  \begin{figure}
	\centering
	
		{}\includegraphics[width=0.43\textwidth]{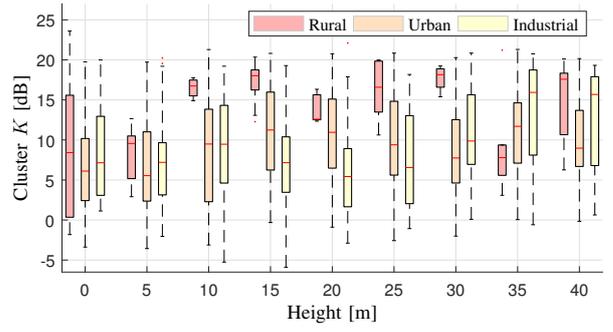}{
		\psfrag{K [dB]}[c][c][0.7]{Cluster $K$ [dB]}
		\psfrag{Height [m]}[c][c][0.7]{Height [m]}
		\psfrag{Urban}[l][l][0.6]{Urban}
		\psfrag{Rural}[l][l][0.6]{Rural}
		\psfrag{Industrialo}[l][l][0.6]{Industrial}
		}
		\caption{Cluster power ratios $K$ at different heights in the three scenarios. \label{fig:cluster_K}}

\end{figure}

% \begin{figure}
% \centering
% {}\includegraphics[width=0.43\textwidth]{cluster_number}{
% \psfrag{cluster number}[c][c][0.7]{Cluster number}
% \psfrag{Height [m]}[c][c][0.7]{Height [m]}
% \psfrag{Urban}[l][l][0.6]{Urban}
% \psfrag{Rural}[l][l][0.6]{Rural}
% \psfrag{Industrialo}[l][l][0.6]{Industrial}
% }
% \caption{The numbers of clusters $C$ for multiple-clusters channels at different heights in the three scenarios. \label{fig:cluster_number}}
% \end{figure}
%
%
% \begin{figure}
% \centering
% {}\includegraphics[width=0.43\textwidth]{cluster_k}{
% \psfrag{K [dB]}[c][c][0.7]{Cluster $K$ [dB]}
% \psfrag{Height [m]}[c][c][0.7]{Height [m]}
% \psfrag{Urban}[l][l][0.6]{Urban}
% \psfrag{Rural}[l][l][0.6]{Rural}
% \psfrag{Industrialo}[l][l][0.6]{Industrial}
% }
% \caption{Cluster power ratios $K$ at different heights in the three scenarios. \label{fig:cluster_K}}
% \end{figure}

\begin{figure}
\centering
{
	{
{}\includegraphics[width=0.45\textwidth]{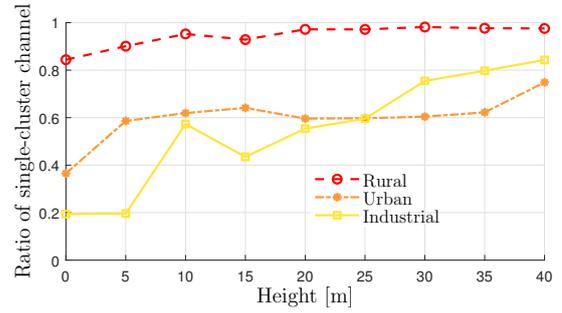}
}}{
\psfrag{Height [m]}[c][c][0.7]{Height [m]}
\psfrag{ratio of channels with one cluster}[c][c][0.7]{Ratio of single-cluster channel}
\psfrag{data1}[l][l][0.6]{Rural}
\psfrag{data2}[l][l][0.6]{Urban}
\psfrag{data3}[l][l][0.6]{Industrial}
}
\caption{Ratios of single-cluster channels at different heights in the three scenarios. \label{fig:one_cluster_ratio}}
\end{figure}

For a channel $\hat \bsTheta$, $C$ clusters are identified and sorted with power-descending order. The cluster power ratio $K$ is defined as
\begin{equation}
\begin{aligned}
K= \frac{p_1}{\sum_{c=2}^{C} p_c},
\end{aligned}\label{eq:k}
\end{equation} %with
\begin{equation}
\begin{aligned}
p_c= \sum_{\ell \in \mathcal{L}_c} |\alpha_\ell|^2
\end{aligned}\label{eq:clsuter_power}.
\end{equation} $K$ demonstrates the importance of the dominant cluster in the channel. A very high $K$ means that the other clusters can be negligible for communications, whereas a small $K$ means that the other clusters can be potentially exploited for communications as they contain non-negligible power. Note that $C$ can be 1 for some channels. Thus, single-cluster channels are not considered when calculating $K$. \frefp{fig:cluster_number} and \ref{fig:cluster_K} illustrate the number of clusters and the cluster power ratios $K$ for multiple-clusters channels at different heights in the three scenarios. The corresponding ratios of single-cluster channels, defined as the ratio of the number of the channels with only one cluster to the number of all the channels, are also illustrated in \fref{fig:one_cluster_ratio}.\footnote{\addnewfr{Note that a single-path channel is a single-cluster channel. The reverse is not necessarily true.}} It can be observed from \fref{fig:cluster_number} that the number of clusters becomes smaller with increasing heights for all the three scenarios. The number of clusters in the industrial scenario is the largest among the three scenarios, which is also consistent with its largest composite spreads as illustrated in \fref{fig:comspreads}. As for
$K$, it can be observed from \fref{fig:cluster_K} that
it is not large at lower heights, i.e. 0 and 5\,m, for all the scenarios. In the rural scenario, $K$ basically increases with increasing heights. However,  $K$ becomes much lower at 35\,m as an exception. We have illustrated the type of \acp{APDP} in \fref{fig:rural_pdp_b}, which is probably because the power of the \ac{LoS} cluster decreases due to the downtilt \ac{BS} antenna pattern, and the reflected component(s) from a certain scatterer that can be detected by the \ac{UAV}-\ac{UE} at this height have non-negligible power(s). For the urban and industrial scenarios, $K$ has similar behaviours. That is, it increases, then decreases and increases again with increasing heights. We postulate this is mainly because at middle heights, e.g. around 20-25\,m in the industrial scenario, building roofs or sidewalls in both scenarios can lead to clusters with relatively high powers. Specially, the median of $K$ can be lower as around 5\,dB in the industrial scenario at the height of 20\,m. Nevertheless, it can be observed from \fref{fig:one_cluster_ratio} that the ratio of single cluster channel tends to increase with increasing heights for all the three scenarios (where the rural scenario has the largest ratios). This indicates that most of the channels at a higher height in all the three scenarios are with smaller spreads.

\subsection{Intra-cluster delay spread $\sigma_{\tau_c}$ and azimuth spread $\sigma_{\phi_c}$}
\begin{figure}
\begin{center}
\subfigure[]{{}\includegraphics[width=0.42\textwidth]{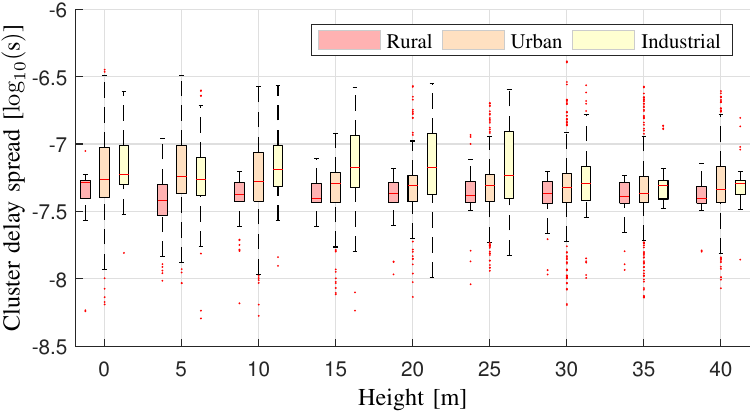}{
\psfrag{cludelayspread [log10(s)]}[c][c][0.7]{Cluster delay spread [$\log_{10}(\text{s})$]}
\psfrag{Height [m]}[c][c][0.7]{Height [m]}
\psfrag{Urban}[l][l][0.6]{Urban}
\psfrag{Rural}[l][l][0.6]{Rural}
\psfrag{Industrialo}[l][l][0.6]{Industrial}
}\label{fig:cluster_delay_spread}}
\subfigure[]{{}\includegraphics[width=0.42\textwidth]{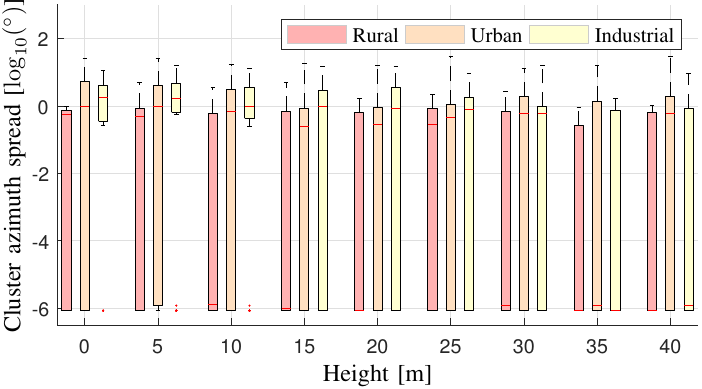}{
\psfrag{Cluazimuthspread [log10(deg)]}[c][c][0.7]{Cluster azimuth spread [$\log_{10}(^\circ)$]}
\psfrag{Height [m]}[c][c][0.7]{Height [m]}
\psfrag{Urban}[l][l][0.6]{Urban}
\psfrag{Rural}[l][l][0.6]{Rural}
\psfrag{Industrialo}[l][l][0.6]{Industrial}
}\label{fig:cluster_azimuth_spread}}
\end{center}
\caption{The intra-cluster delay spreads and azimuth spreads observed in the three scenarios. (a) Intra-cluster delay spreads $\sigma_{\tau_c}$. (b) Intra-cluster azimuth spreads $\sigma_{\phi_c}$. \label{fig:clusterspreads}}
\end{figure}

\begin{figure}
\centering
{
	{
{}\includegraphics[width=0.45\textwidth]{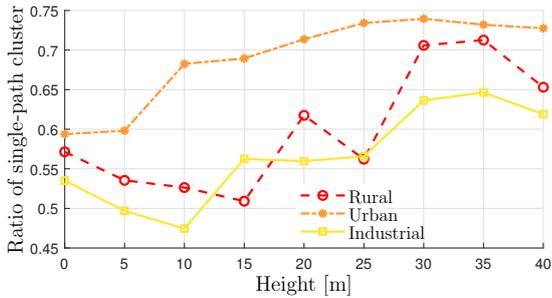}
}}{
\psfrag{Height [m]}[c][c][0.7]{Height [m]}
\psfrag{single-path-cluster ratio}[c][c][0.7]{Ratio of single-path cluster}
\psfrag{data1}[l][l][0.6]{Rural}
\psfrag{data2}[l][l][0.6]{Urban}
\psfrag{data3}[l][l][0.6]{Industrial}
}
\caption{Ratios of single-path clusters at different heights in the three scenarios.  \label{fig:ratio_one_path_cluster}}
\end{figure}

The intra-cluster delay spread and azimuth spread of an identified cluster are calculated using \eqref{eq:delayspread} and \eqref{eq:azimuthspread}, respectively, with $\mathcal L$ replaced with $\mathcal L_c$. Similarly, clusters with a single path are excluded. \fref{fig:clusterspreads} illustrates the observed intra-cluster delay and azimuth spreads, both indicated in logarithm scales, for the three scenarios. The ratios of single-path clusters, defined as the ratio of the number of the clusters with only one path to the number of all the clusters, are illustrated in \fref{fig:ratio_one_path_cluster}. It can be observed from \fref{fig:clusterspreads} that the intra-cluster spreads are the smallest in the rural scenario, and they have weak dependence with heights. We conjecture this is because the rural scenario is very open, and most of the clusters are attributed to the \ac{LoS} links. In the urban scenario, the intra-cluster spreads basically decrease with increasing heights. However, in the industrial scenario, it can be observed that the intra-cluster delay spreads can become larger at a certain height, e.g. at 25\,m, although the overall trend is to decrease with increasing heights. We postulate this is mainly because the urban scenario and the industrial scenario have different types of buildings. In the industrial scenarios, several tall building with round sidewalls can exist in a certain region together with some lower building roofs, which can result in a cluster of path components with a larger intra-cluster delay spread. However, in the urban scenario, there are no round and tall concrete-elements, and houses roofs are usually spire-alike with their slopes towards different directions. Thus, clusters, each caused by a single roof, with smaller spreads both in delay and azimuth are more possible to occur. This can also be inferred from \fref{fig:ratio_one_path_cluster} that the ratio of single-path clusters in the urban scenario is larger than that in the industrial scenario in most cases. Besides, the single-path-cluster ratios basically increase with increasing heights in all the three scenarios. It is also worth noting that the rural scenario do not have the highest ratios as those illustrated in
\fref{fig:ratio_one_path_com} or \fref{fig:one_cluster_ratio}. This is because a channel in the urban and industrial scenarios is more possible to have multiple clusters, among which several single-path clusters can exist.

\subsection{Inter-cluster delay offsets $\tau_{\text o}$, azimuth offsets $\phi_{\text o}$ and power offsets $p_\text{o}$}

The inter-cluster delay/azimuth offsets are calculated as the differences between the mean delays/azimuths of the non-dominant clusters and the mean delay/azimuth of the dominant cluster of a channel. They indicate the separateness of the multiple clusters in terms of delay and azimuth. For example, a large azimuth offset means that the spatial diversity may be exploited using multiple horn antennas at the \ac{UAV}-\ac{UE} side.

\begin{figure}%[!tbp]
  \centering
		{}\includegraphics[width=0.43\textwidth]{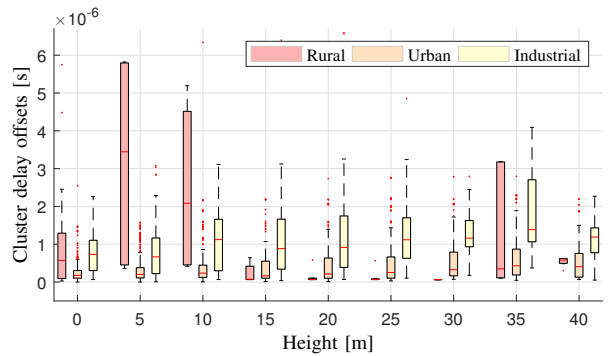}{
		\psfrag{Delay offsets [us]}[c][c][0.7]{Cluster delay offsets [s]}
		\psfrag{Height [m]}[c][c][0.7]{Height [m]}
		\psfrag{Urban}[l][l][0.6]{Urban}
		\psfrag{Rural}[l][l][0.6]{Rural}
		\psfrag{Industrialo}[l][l][0.6]{Industrial}
		}
		\caption{The inter-cluster delay offsets of multiple-clusters channels at different heights in the three scenarios. \label{fig:delay_offsets}}
\end{figure}
\begin{figure}
	\centering
	
	{}\includegraphics[width=0.43\textwidth]{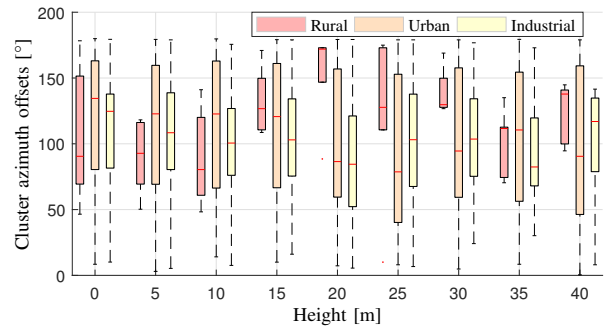}{
	\psfrag{Azimuth offsets [deg]}[c][c][0.7]{Cluster azimuth offsets [$^\circ$]}
	\psfrag{Height [m]}[c][c][0.7]{Height [m]}
	\psfrag{Urban}[l][l][0.6]{Urban}
	\psfrag{Rural}[l][l][0.6]{Rural}
	\psfrag{Industrialo}[l][l][0.6]{Industrial}
	}
	\caption{The inter-cluster azimuth offsets of multiple-clusters channels at different heights in the three scenarios. \label{fig:azimuth_offset}}

\end{figure}

% \begin{figure}
% \centering
% {}\includegraphics*[width=0.43\textwidth]{delay_offsets}{
% \psfrag{Delay offsets [us]}[c][c][0.7]{Cluster delay offsets [s]}
% \psfrag{Height [m]}[c][c][0.7]{Height [m]}
% \psfrag{Urban}[l][l][0.6]{Urban}
% \psfrag{Rural}[l][l][0.6]{Rural}
% \psfrag{Industrialo}[l][l][0.6]{Industrial}
% }
% \caption{The inter-cluster delay offsets of multiple-clusters channels at different heights in the three scenarios. \label{fig:delay_offsets}}
% \end{figure}
%
% \begin{figure}
% \centering
% {}\includegraphics[width=0.43\textwidth]{azimuth_offsets}{
% \psfrag{Azimuth offsets [deg]}[c][c][0.7]{Cluster azimuth offsets [$^\circ$]}
% \psfrag{Height [m]}[c][c][0.7]{Height [m]}
% \psfrag{Urban}[l][l][0.6]{Urban}
% \psfrag{Rural}[l][l][0.6]{Rural}
% \psfrag{Industrialo}[l][l][0.6]{Industrial}
% }
% \caption{The inter-cluster azimuth offsets of multiple-clusters channels at different heights in the three scenarios. \label{fig:azimuth_offset}}
% \end{figure}

\fref{fig:delay_offsets} illustrates the delay offsets at different heights in the three scenarios. Note that they are only calculated for the channels with multiple clusters. It can be observed from \fref{fig:delay_offsets} that in both the industrial scenarios and the urban scenarios, delay offsets become larger with increasing heights. This is reasonable since with a higher height, it is more possible that farther objects can also contribute to the received signals. However, it is obvious that in the urban scenario the delay offsets do not increase as fast as that observed in the industrial scenario. We postulate this is also due to the different types of buildings in the two scenarios. With very tall concrete elements in the industrial scenario, path components with much farther delays can be observed with higher heights. As for the rural scenario, delay offsets are quite small at most of the heights as there is no building in a close proximity in the open field, except that in some cases path components from very-far objects can lead to very large delay offsets. \fref{fig:azimuth_offset} illustrates the azimuth offsets at different heights in the three scenarios. It can be observed that there are no certain trends can be observed with respect to heights. Overall, the deviations of the azimuth offsets in the urban scenario are the largest due to its high density of buildings therein. Moreover, the azimuth offsets are usually large as above 60$^\circ$, which indicates the multiple clusters are well separated in the azimuth domain.

\begin{figure}
\centering
{}\includegraphics[width=0.43\textwidth]{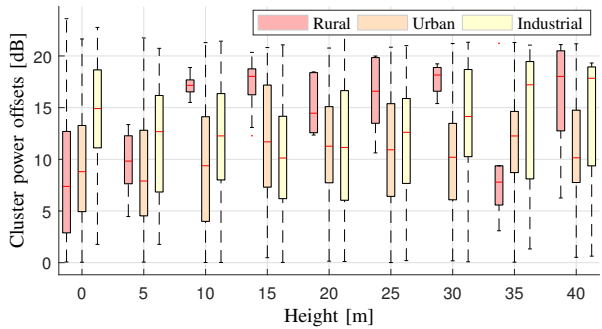}{
\psfrag{Power offsets [dB]}[c][c][0.7]{Cluster power offsets [dB]}
\psfrag{Height [m]}[c][c][0.7]{Height [m]}
\psfrag{Urban}[l][l][0.6]{Urban}
\psfrag{Rural}[l][l][0.6]{Rural}
\psfrag{Industrialo}[l][l][0.6]{Industrial}
}
\caption{The inter-cluster power offsets of multiple-clusters channels at different heights in the three scenarios. \label{fig:power_offset}}
\end{figure}

We also inspected the power decay behaviours of clusters with respect to their delays and do not find certain dependences, e.g. exponential decay, between cluster powers and cluster delays. We postulate this is due to the fact that the powers of clusters not only related to the propagation distances but also the radiation patterns of the cellular \acp{BS}. It is probably random that a cluster may be caused by the main lobe or sidelobes of a \ac{BS}. Nevertheless, we calculated the inter-cluster power offsets as the difference between the power in dB of the dominant cluster and the powers in dB of the other non-dominant clusters. \fref{fig:power_offset} illustrates the power offsets at different heights in the three scenarios. The consistency between \fref{fig:power_offset} and \fref{fig:cluster_K} can be well observed.  For example, in the rural scenario, the power offsets are quite high at above 10\,m (except 35\,m), which results in high $K$'s at these heights in \fref{fig:cluster_K}. In the industrial scenario, the power offsets at 20\,m have the lowest values among all the heights, which corresponds to the lowest cluster
$K$, as illustrated in \fref{fig:cluster_K}, for the industrial scenario.

% without medians, etc.

\begin{table*}
\centering
\caption{Statistics extracted for the spatial channels at different heights in the three scenarios. The three lines of parameter values from top to bottom in each cell correspond to the rural, urban and industrial scenarios, respectively. $r_1$, $r_2$ and $r_3$ represent the ratios of single-path channels, single-cluster channels and single-path clusters, respectively. \addnewfr{Due to space limitation, we do not include the 25th-percentile, 75th-percentile, minimum and maximum of each parameter sample, which can be found in the ``boxplot'' figures, e.g., \fref{fig:cluster_K} for cluster K.}}
\label{tab:statistics}\scalebox{0.85}{
\begin{tabular}{c|c|c|c|c|c}
                                                                & \multicolumn{5}{c}{Channel parameter values: (mean, std.)}                                                                                                                                                                                                                                                                                                                                                                                                                            \\
\hline
\begin{tabular}[c]{@{}c@{}}Height \\ $[\text{m}]$ \end{tabular} & \begin{tabular}[c]{@{}c@{}}Ratios\\($r_1$; $r_2$; $r_3$) \end{tabular}                                   & \begin{tabular}[c]{@{}c@{}}Composite delay\\spreads [$\log_{10}(\text{s})$] \end{tabular} & \begin{tabular}[c]{@{}c@{}}Composite azimuth \\spreads [$\log_{10}(^\circ)$] \end{tabular} & Cluster number                                                                & Cluster $K$ [dB]                                                                 \\
\hline
0                                                               & \begin{tabular}[c]{@{}c@{}} (0.45; 0.84; 0.57)\\ (0.17; 0.36; 0.59)\\ (0.06; 0.19; 0.54)\\ \end{tabular} & \begin{tabular}[c]{@{}c@{}}(-7.24, 0.51)\\(-7.12, 0.36)\\(-6.74, 0.40) \end{tabular}                       & \begin{tabular}[c]{@{}c@{}}(-1.0, 2.9)\\(1.1, 1.8)\\(0.6, 2.3)\end{tabular}                & \begin{tabular}[c]{@{}c@{}}(2.3, 0.5)\\(2.7, 1.1)\\(4.3, 1.9)\\ \end{tabular} & \begin{tabular}[c]{@{}c@{}}(9.0, 8.4)\\(6.5, 5.1)\\(8.6, 6.3)\\\end{tabular}     \\
\hline
5                                                               & \begin{tabular}[c]{@{}c@{}} (0.46; 0.90; 0.54)\\ (0.26; 0.59; 0.60)\\ (0.10; 0.20; 0.50)\\ \end{tabular} & \begin{tabular}[c]{@{}c@{}}(-7.23, 0.64)\\(-7.10, 0.43)\\(-6.75, 0.47) \end{tabular}                       & \begin{tabular}[c]{@{}c@{}}(-1.6, 3.1)\\(0.6, 2.2)\\(0.9, 2.2)\end{tabular}                & \begin{tabular}[c]{@{}c@{}}(2.5, 0.5)\\(2.7, 1.2)\\(3.7, 1.3)\\ \end{tabular} & \begin{tabular}[c]{@{}c@{}}(8.4, 3.3)\\(6.8, 5.8)\\(7.0, 5.6)\\\end{tabular}     \\
\hline
10                                                              & \begin{tabular}[c]{@{}c@{}} (0.48; 0.95; 0.53)\\ (0.38; 0.62; 0.68)\\ (0.24; 0.57; 0.47)\\ \end{tabular} & \begin{tabular}[c]{@{}c@{}}(-7.33, 0.31)\\(-7.16, 0.44)\\(-7.08, 1.78) \end{tabular}                       & \begin{tabular}[c]{@{}c@{}}(-2.9, 3.1)\\(0.1, 2.7)\\(0.1, 2.3)\end{tabular}                & \begin{tabular}[c]{@{}c@{}}(2.2, 0.4)\\(2.4, 0.8)\\(3.9, 1.9)\\ \end{tabular} & \begin{tabular}[c]{@{}c@{}}(16.5, 1.1)\\(8.6, 6.4)\\(8.4, 7.1)\\\end{tabular}    \\
\hline
15                                                              & \begin{tabular}[c]{@{}c@{}} (0.43; 0.93; 0.51)\\ (0.48; 0.64; 0.69)\\ (0.27; 0.43; 0.56)\\ \end{tabular} & \begin{tabular}[c]{@{}c@{}}(-7.43, 0.23)\\(-7.29, 0.37)\\(-7.22, 2.81) \end{tabular}                        & \begin{tabular}[c]{@{}c@{}}(-2.6, 3.2)\\(-1.0, 3.1)\\(0.4, 2.5)\end{tabular}               & \begin{tabular}[c]{@{}c@{}}(2.0, 0.0)\\(2.2, 0.5)\\(3.7, 2.0)\\ \end{tabular} & \begin{tabular}[c]{@{}c@{}}(17.3, 2.6)\\(11.0, 5.5)\\(7.2, 6.4)\\\end{tabular}   \\
\hline
20                                                              & \begin{tabular}[c]{@{}c@{}} (0.58; 0.97; 0.62)\\ (0.55; 0.60; 0.71)\\ (0.37; 0.55; 0.56)\\ \end{tabular} & \begin{tabular}[c]{@{}c@{}}(-7.40, 0.18)\\(-7.24, 0.43)\\(-7.09, 2.44) \end{tabular}                       & \begin{tabular}[c]{@{}c@{}}(-3.6, 3.0)\\(-1.1, 3.2)\\(-0.6, 3.2)\\\end{tabular}            & \begin{tabular}[c]{@{}c@{}}(2.2, 0.4)\\(2.2, 0.5)\\(3.7, 1.5)\\ \end{tabular} & \begin{tabular}[c]{@{}c@{}}(13.9, 1.9)\\(10.7, 5.3)\\(6.0, 6.0)\\\end{tabular}   \\
\hline
25                                                              & \begin{tabular}[c]{@{}c@{}} (0.52; 0.97; 0.56)\\ (0.57; 0.60; 0.73)\\ (0.30; 0.60; 0.57)\\ \end{tabular} & \begin{tabular}[c]{@{}c@{}}(-7.39, 0.21)\\(-7.20, 0.42)\\(-7.21, 2.61) \end{tabular}                       & \begin{tabular}[c]{@{}c@{}}(-2.7, 3.1)\\(-1.1, 3.1)\\(-0.5, 3.0)\\\end{tabular}            & \begin{tabular}[c]{@{}c@{}}(2.0, 0.0)\\(2.4, 0.7)\\(3.5, 1.8)\\ \end{tabular} & \begin{tabular}[c]{@{}c@{}}(16.2, 4.0)\\(9.7, 6.1)\\(7.8, 6.1)\\\end{tabular}    \\
\hline
30                                                              & \begin{tabular}[c]{@{}c@{}} (0.68; 0.98; 0.71)\\ (0.62; 0.60; 0.74)\\ (0.58; 0.75; 0.64)\\ \end{tabular} & \begin{tabular}[c]{@{}c@{}}(-7.44, 0.23)\\(-7.18, 0.83)\\(-6.90, 0.50) \end{tabular}                       & \begin{tabular}[c]{@{}c@{}}(-3.3, 3.0)\\(-1.2, 3.2)\\(-1.6, 3.5)\\\end{tabular}            & \begin{tabular}[c]{@{}c@{}}(2.0, 0.0)\\(2.3, 0.6)\\(2.8, 1.4)\\ \end{tabular} & \begin{tabular}[c]{@{}c@{}}(17.7, 1.7)\\(8.7, 5.2)\\(10.9, 6.4)\\\end{tabular}   \\
\hline
35                                                              & \begin{tabular}[c]{@{}c@{}} (0.69; 0.98; 0.71)\\ (0.62; 0.62; 0.73)\\ (0.51; 0.80; 0.65)\\ \end{tabular} & \begin{tabular}[c]{@{}c@{}}(-7.35, 0.26)\\(-7.20, 0.44)\\(-7.01, 0.47) \end{tabular}                       & \begin{tabular}[c]{@{}c@{}}(-4.1, 2.9)\\(-1.6, 3.3)\\(-2.6, 3.5)\\\end{tabular}            & \begin{tabular}[c]{@{}c@{}}(2.0, 0.0)\\(2.3, 0.7)\\(2.3, 0.4)\\ \end{tabular} & \begin{tabular}[c]{@{}c@{}}(9.1, 6.4)\\(10.8, 4.8)\\(13.6, 6.8)\\\end{tabular}   \\
\hline
40                                                              & \begin{tabular}[c]{@{}c@{}}(0.64; 0.98; 0.65)\\(0.64; 0.75; 0.73)\\(0.50; 0.84; 0.62) \end{tabular}      & \begin{tabular}[c]{@{}c@{}}(-7.37, 0.15)\\(-7.21, 0.37)\\(-7.10, 0.36) \end{tabular}                       & \begin{tabular}[c]{@{}c@{}}(-3.3, 3.1)\\(-1.7, 3.3)\\(-2.1, 3.3)\\\end{tabular}            & \begin{tabular}[c]{@{}c@{}}(2.1, 0.4)\\(2.3, 0.8)\\(2.3, 0.5)\\ \end{tabular} & \begin{tabular}[c]{@{}c@{}}(15.1, 5.4)\\(9.7, 5.0)\\(12.2, 6.6)\\\end{tabular}   \\
\hline\hline
\begin{tabular}[c]{@{}c@{}}Height \\ $[\text{m}]$ \end{tabular} & \begin{tabular}[c]{@{}c@{}}Cluster azimuth \\spreads [$\log_{10}(^\circ)$] \end{tabular}                 & \begin{tabular}[c]{@{}c@{}}Cluster delay \\spreads [$\log_{10}(\text{s})$] \end{tabular}                   & \begin{tabular}[c]{@{}c@{}}Cluster delay \\offsets [$\log_{10}(\text{s})$] \end{tabular}   & \begin{tabular}[c]{@{}c@{}}Cluster azimuth \\offsets [$^\circ$] \end{tabular} & \begin{tabular}[c]{@{}c@{}}Cluster power \\offsets [dB] \end{tabular}            \\
\hline
0                                                               & \begin{tabular}[c]{@{}c@{}}(-2.1, 2.7)\\(-1.6, 3.0)\\(-1.0, 2.7)\\\end{tabular}                          & \begin{tabular}[c]{@{}c@{}}(-7.38, 0.24)\\(-7.20, 0.31)\\(-7.17, 0.22)\\\end{tabular}                      & \begin{tabular}[c]{@{}c@{}}(-5.90, -5.73)\\(-6.60, -6.59)\\(-6.15, -6.35)\\ \end{tabular}  & \begin{tabular}[c]{@{}c@{}}(103, 46)\\(119, 49)\\(113, 42)\\ \end{tabular} & \begin{tabular}[c]{@{}c@{}}(8.4, 7.1)\\(9.1, 5.3)\\(14.5, 5.4)\\\end{tabular}    \\
\hline
5                                                               & \begin{tabular}[c]{@{}c@{}}(-2.3, 3.0)\\(-1.3, 2.8)\\(-1.0, 2.7)\\\end{tabular}                          & \begin{tabular}[c]{@{}c@{}}(-7.47, 0.23)\\(-7.18, 0.28)\\(-7.25, 0.30)\\\end{tabular}                      & \begin{tabular}[c]{@{}c@{}}(-5.49, -5.60)\\(-6.49, -6.47)\\(-6.06, -6.11)\\ \end{tabular}  & \begin{tabular}[c]{@{}c@{}}(91, 25)\\(112, 51)\\(106, 41)\\ \end{tabular} & \begin{tabular}[c]{@{}c@{}}(9.6, 2.8)\\(8.7, 5.3)\\(11.6, 5.4)\\\end{tabular}    \\
\hline
10                                                              & \begin{tabular}[c]{@{}c@{}}(-3.2, 2.9)\\(-2.1, 3.1)\\(-1.0, 2.5)\\\end{tabular}                          & \begin{tabular}[c]{@{}c@{}}(-7.40, 0.18)\\(-7.24, 0.29)\\(-7.30, 1.36)\\\end{tabular}                      & \begin{tabular}[c]{@{}c@{}}(-5.63, -5.67)\\(-6.41, -6.28)\\(-5.94, -6.04)\\ \end{tabular}  & \begin{tabular}[c]{@{}c@{}}(91, 35)\\(114, 52)\\(104, 36)\\ \end{tabular} & \begin{tabular}[c]{@{}c@{}}(17.2, 1.1)\\(9.3, 5.8)\\(11.5, 5.7)\\\end{tabular}   \\
\hline
15                                                              & \begin{tabular}[c]{@{}c@{}}(-3.4, 3.0)\\(-2.9, 3.0)\\(-1.7, 3.1)\\\end{tabular}                          & \begin{tabular}[c]{@{}c@{}}(-7.37, 0.10)\\(-7.33, 0.21)\\(-7.50, 2.20)\\\end{tabular}                      & \begin{tabular}[c]{@{}c@{}}(-6.67, -6.61)\\(-6.43, -6.34)\\(-5.93, -5.92)\\ \end{tabular}  & \begin{tabular}[c]{@{}c@{}}(132, 25)\\(110, 53)\\(105, 41)\\ \end{tabular} & \begin{tabular}[c]{@{}c@{}}(17.3, 2.6)\\(11.8, 5.5)\\(10.2, 5.2)\\\end{tabular}  \\
\hline
20                                                              & \begin{tabular}[c]{@{}c@{}}(-3.9, 2.9)\\(-2.8, 3.1)\\(-1.8, 3.1)\\\end{tabular}                          & \begin{tabular}[c]{@{}c@{}}(-7.38, 0.14)\\(-7.32, 0.21)\\(-7.39, 1.86)\\\end{tabular}                      & \begin{tabular}[c]{@{}c@{}}(-6.80, -6.68)\\(-6.33, -6.25)\\(-5.89, -5.87)\\ \end{tabular}  & \begin{tabular}[c]{@{}c@{}}(154, 34)\\(100, 53)\\(86, 44)\\ \end{tabular} & \begin{tabular}[c]{@{}c@{}}(15.1, 3.0)\\(11.2, 5.0)\\(11.1, 5.9)\\\end{tabular}  \\
\hline
25                                                              & \begin{tabular}[c]{@{}c@{}}(-3.0, 3.0)\\(-2.7, 3.1)\\(-2.0, 3.0)\\\end{tabular}                          & \begin{tabular}[c]{@{}c@{}}(-7.36, 0.15)\\(-7.32, 0.22)\\(-7.51, 2.24)\\\end{tabular}                      & \begin{tabular}[c]{@{}c@{}}(-6.82, -6.69)\\(-6.27, -6.19)\\(-5.89, -6.05)\\ \end{tabular}  & \begin{tabular}[c]{@{}c@{}}(121, 60)\\(96, 55)\\(103, 48)\\ \end{tabular} & \begin{tabular}[c]{@{}c@{}}(16.2, 4.0)\\(10.6, 5.8)\\(11.6, 5.3)\\\end{tabular}  \\
\hline
30                                                              & \begin{tabular}[c]{@{}c@{}}(-3.5, 2.9)\\(-2.6, 3.2)\\(-2.4, 3.1)\\\end{tabular}                          & \begin{tabular}[c]{@{}c@{}}(-7.39, 0.17)\\(-7.36, 0.82)\\(-7.26, 0.33)\\\end{tabular}                      & \begin{tabular}[c]{@{}c@{}}(-7.20, -8.33)\\(-6.23, -6.24)\\(-5.90, -6.20)\\ \end{tabular}  & \begin{tabular}[c]{@{}c@{}}(139, 20)\\(104, 54)\\(108, 36)\\ \end{tabular} & \begin{tabular}[c]{@{}c@{}}(17.7, 1.7)\\(9.8, 5.0)\\(13.8, 5.6)\\\end{tabular}   \\
\hline
35                                                              & \begin{tabular}[c]{@{}c@{}}(-4.5, 2.6)\\(-2.9, 3.2)\\(-4.0, 2.9)\\\end{tabular}                          & \begin{tabular}[c]{@{}c@{}}(-7.39, 0.13)\\(-7.35, 0.27)\\(-7.28, 0.17)\\\end{tabular}                      & \begin{tabular}[c]{@{}c@{}}(-5.92, -5.81)\\(-6.23, -6.25)\\(-5.76, -5.99)\\ \end{tabular}  & \begin{tabular}[c]{@{}c@{}}(103, 25)\\(103, 55)\\(91, 33)\\ \end{tabular} & \begin{tabular}[c]{@{}c@{}}(9.1, 6.4)\\(11.6, 4.3)\\(14.4, 6.8)\\\end{tabular}   \\
\hline
40                                                              & \begin{tabular}[c]{@{}c@{}}(-3.4, 2.9)\\(-2.6, 3.2)\\(-3.5, 3.1)\\\end{tabular}                          & \begin{tabular}[c]{@{}c@{}}(-7.39, 0.09)\\(-7.28, 0.25)\\(-7.29, 0.18)\\\end{tabular}                      & \begin{tabular}[c]{@{}c@{}}(-6.27, -6.96)\\(-6.28, -6.31)\\(-5.90, -6.21)\\ \end{tabular}  & \begin{tabular}[c]{@{}c@{}}(125, 22)\\(99, 57)\\(97, 49)\\ \end{tabular} & \begin{tabular}[c]{@{}c@{}}(16.2, 5.7)\\(10.8, 4.9)\\(13.7, 6.5)\\\end{tabular}  \\
\hline
\end{tabular}
}

\end{table*}

\section{Conclusions} \label{section:conclusion}
In this contribution, a measurement campaign was conducted using a 16-elements \acf{UCA} to collect the downlink signals from the \acf{LTE} networks at the heights from 0 to 40\,m in the rural, urban and industrial scenarios. High-resolution channel parameters were estimated from the extracted \acfp{CIR}, and comprehensive composite and cluster-level characteristics of the \acf{A2G} spatial channels were investigated. Generally, the \ac{A2G} channels become less complicated when the height is increased in all the three scenarios, i.e. with smaller composite spreads, cluster number and cluster spreads, larger cluster $K$, etc. observed. The rural scenario has the least-complicated channels due to its openness, most of which are with only one cluster. It can be considered that the rural channels become \ac{LoS} dominant with increasing heights. Nevertheless, it is still possible that for a rural channel at a higher height, an additional cluster with a relatively high power, a long delay and a large angle offset can exist. In both the urban and the industrial scenarios, the channels at low heights, e.g. 0 and 5\,m, are observed with more clusters than that observed in the rural scenario, caused by the higher building densities. Moreover, it is interesting to find that properly higher heights in both scenarios can result in more complicated channels though the height is increased. For example, at 20\,m in the industrial scenario, around 40\% channels have multiple clusters, and the cluster $K$ can be 5\,dB and even lower. We postulate this is mainly because of contributions from the building roofs and/or sidewalls in both scenarios. The obtained results show that assuming the most of the channels are pure or close to \acf{LoS} even at higher heights is non-realistic, at least in the complex scenarios such as urban and industrial scenarios with rich number of buildings, for technology verification and performance evaluation. Furthermore, clusters from farther-away scatterers emerge with increasing heights in all the scenarios, since the cluster delay offsets are observed to be larger. The large cluster azimuth offsets, with medians more than $60^\circ$, observed in all the three scenarios also demonstrates that multiple clusters are well separated in the angular domain. This means that multiple-antennas techniques are promising for cellular-connected \acp{UAV}. Table\,\ref{tab:statistics} summarizes all the extracted spatial channel characteristics at different heights in all the three scenarios, which readers can refer to for specific modeling-parameter values. The spatial characteristics investigated in this contribution are important for understanding the low-altitude \ac{A2G} channels to enable the advanced communications for both cellular-connected \acfp{UAV} and ground \acfp{UE} in \acs{5G} and beyond networks. Future work will exploit the established channel models to facilitate the proposal and evaluation of solutions for \ac{UAV} connectivity in cellular networks. {As a final note, further measurements and investigations are also needed to improve the modeling parameters set to involve the elevation characteristics and spatial consistency, i.e., how the channel evolves with the aerial vehicle moving.}

\setlength{\itemsep}{0em}
\renewcommand{\baselinestretch}{1.08}
\patchcmd{\thebibliography}
  {\settowidth}
  {\setlength{\parsep}{0pt}\setlength{\itemsep}{0pt plus 0.1pt}\settowidth}
  {}{}

\bibliographystyle{IEEEtran}
% \bibliography{uav_spatial_channel_modeling.bbl}
\bibliography{IEEEabrv,uav_bib_xuesong,uav_bib_xose,CvBibliographicDatabases/Eu/Bibliografia_CV_33540678P_en-Shortened}

\begin{thebibliography}{10}
\providecommand{\url}[1]{#1}
\csname url@samestyle\endcsname
\providecommand{\newblock}{\relax}
\providecommand{\bibinfo}[2]{#2}
\providecommand{\BIBentrySTDinterwordspacing}{\spaceskip=0pt\relax}
\providecommand{\BIBentryALTinterwordstretchfactor}{4}
\providecommand{\BIBentryALTinterwordspacing}{\spaceskip=\fontdimen2\font plus
\BIBentryALTinterwordstretchfactor\fontdimen3\font minus
  \fontdimen4\font\relax}
\providecommand{\BIBforeignlanguage}[2]{{%
\expandafter\ifx\csname l@#1\endcsname\relax
\typeout{** WARNING: IEEEtran.bst: No hyphenation pattern has been}%
\typeout{** loaded for the language `#1'. Using the pattern for}%
\typeout{** the default language instead.}%
\else
\language=\csname l@#1\endcsname
\fi
#2}}
\providecommand{\BIBdecl}{\relax}
\BIBdecl

\bibitem{hayat2016survey}
S.~Hayat, E.~Yanmaz, and R.~Muzaffar, ``Survey on unmanned aerial vehicle
  networks for civil applications: A communications viewpoint,'' \emph{{IEEE}
  Commun. Surveys Tuts.}, vol.~18, no.~4, pp. 2624--2661, 2016.

\bibitem{kumar2018unmanned}
N.~{Kumar}, D.~{Puthal}, T.~{Theocharides}, and S.~P. {Mohanty}, ``Unmanned
  aerial vehicles in consumer applications: New applications in current and
  future smart environments,'' \emph{IEEE Consum. Electron. Mag.}, vol.~8,
  no.~3, pp. 66--67, 2019.

\bibitem{menouar2017uav}
H.~{Menouar}, I.~{Guvenc}, K.~{Akkaya}, A.~S. {Uluagac}, A.~{Kadri}, and
  A.~{Tuncer}, ``{UAV}-enabled intelligent transportation systems for the smart
  city: Applications and challenges,'' \emph{{IEEE} Commun. Mag.}, vol.~55,
  no.~3, pp. 22--28, 2017.

\bibitem{zeng2016wireless}
Y.~Zeng, R.~Zhang, and T.~J. Lim, ``Wireless communications with unmanned
  aerial vehicles: opportunities and challenges,'' \emph{{IEEE} Commun. Mag.},
  vol.~54, no.~5, pp. 36--42, 2016.

\bibitem{wu2019fundamental}
Q.~{Wu}, L.~{Liu}, and R.~{Zhang}, ``Fundamental trade-offs in communication
  and trajectory design for uav-enabled wireless network,'' \emph{IEEE Wireless
  Communications}, vol.~26, no.~1, pp. 36--44, 2019.

\bibitem{wu2018uav}
Q.~{Wu}, J.~{Xu}, and R.~{Zhang}, ``{UAV}-enabled broadcast channel: Trajectory
  design and capacity characterization,'' in \emph{IEEE International
  Conference on Communications Workshops (ICC Workshops)}, 2018, pp. 1--6.

\bibitem{zeng2017energy}
Y.~{Zeng} and R.~{Zhang}, ``Energy-efficient {UAV} communication with
  trajectory optimization,'' \emph{{IEEE} Trans. Wireless Commun.}, vol.~16,
  no.~6, pp. 3747--3760, 2017.

\bibitem{stocker2017review}
C.~St{\"o}cker, R.~Bennett, F.~Nex, M.~Gerke, and J.~Zevenbergen, ``Review of
  the current state of {UAV} regulations,'' \emph{Remote Sensing}, vol.~9,
  no.~5, 2017.

\bibitem{8470897}
Y.~{Zeng}, J.~{Lyu}, and R.~{Zhang}, ``Cellular-connected {UAV}: Potential,
  challenges, and promising technologies,'' \emph{IEEE Wireless
  Communications}, vol.~26, no.~1, pp. 120--127, 2019.

\bibitem{zeng2019accessing}
Y.~{Zeng}, Q.~{Wu}, and R.~{Zhang}, ``Accessing from the sky: A tutorial on
  {UAV} communications for {5G} and beyond,'' \emph{Proceedings of the IEEE},
  vol. 107, no.~12, pp. 2327--2375, 2019.

\bibitem{3gppuasuav}
3GPP, ``{UAS-UAV},'' 2020, \url{https://www.3gpp.org/uas-uav}.

\bibitem{8581827}
J.~{Stanczak}, D.~{Kozioł}, I.~Z. {Kovács}, J.~{Wigard}, M.~{Wimmer}, and
  R.~{Amorim}, ``Enhanced unmanned aerial vehicle communication support in
  {LTE}-advanced,'' in \emph{IEEE Conf. on Standards for Comm. and Networking},
  2018, pp. 1--6.

\bibitem{nguyen2018how}
H.~C. {Nguyen}, R.~{Amorim}, J.~{Wigard}, I.~Z. {KováCs}, T.~B. {Sørensen},
  and P.~E. {Mogensen}, ``How to ensure reliable connectivity for aerial
  vehicles over cellular networks,'' \emph{IEEE Access}, vol.~6, pp.
  12\,304--12\,317, 2018.

\bibitem{amorim2020enabling}
R.~M. {de Amorim}, J.~{Wigard}, I.~Z. {Kovacs}, T.~B. {Sorensen}, and P.~E.
  {Mogensen}, ``Enabling cellular communication for aerial vehicles: Providing
  reliability for future applications,'' \emph{{IEEE} Veh. Technol. Mag.},
  vol.~15, no.~2, pp. 129--135, 2020.

\bibitem{9082692}
T.~{Izydorczyk}, G.~{Berardinelli}, P.~{Mogensen}, M.~M. {Ginard}, J.~{Wigard},
  and I.~Z. {Kovács}, ``Achieving high {UAV} uplink throughput by using
  beamforming on board,'' \emph{IEEE Access}, vol.~8, pp. 82\,528--82\,538,
  2020.

\bibitem{ArtigoInterferenciaEnUAVs_2019}
X.~Cai, C.~Zhang, J.~Rodr\'iguez-{Pi\~neiro}, X.~Yin, W.~Fan, and G.~F.
  Pedersen, ``Interference modeling for low-height air-to-ground channels in
  live {LTE} networks,'' \emph{IEEE Ant. Wireless Propag. Lett.}, vol.~18,
  no.~10, pp. 2011--2015, 2019.

\bibitem{8369158}
R.~{Amorim}, H.~{Nguyen}, J.~{Wigard}, I.~Z. {Kovács}, T.~B. {Sørensen},
  D.~Z. {Biro}, M.~{Sørensen}, and P.~{Mogensen}, ``Measured uplink
  interference caused by aerial vehicles in {LTE} cellular networks,''
  \emph{IEEE Wireless Commun. Lett.}, vol.~7, no.~6, pp. 958--961, 2018.

\bibitem{cai2020centralized}
X.~Cai, I.~Kovács, J.~Wigard, and P.~Mogensen, ``A centralized and scalable
  uplink power control algorithm in low {SINR}: A case study for {UAV}
  communications,'' arXiv: 2008.06369, 2020.

\bibitem{9099899}
W.~{Mei} and R.~{Zhang}, ``Uplink cooperative interference cancellation for
  cellular-connected {UAV}: A quantize-and-forward approach,'' \emph{IEEE
  Wireless Communications Letters}, vol.~9, no.~9, pp. 1567--1571, 2020.

\bibitem{8763928}
L.~{Liu}, S.~{Zhang}, and R.~{Zhang}, ``Multi-beam {UAV} communication in
  cellular uplink: Cooperative interference cancellation and sum-rate
  maximization,'' \emph{{IEEE} Trans. Wireless Commun.}, vol.~18, no.~10, pp.
  4679--4691, 2019.

\bibitem{UAVResourceAllo}
C.~Fan, ``Joint resource allocation for dynamic cellular-enabled {UAVs}
  communication,'' \emph{IET Communications}, March 2020.

\bibitem{european2013roadmap}
E.~R.~S. Group \emph{et~al.}, ``Roadmap for the integration of civil
  remotely-piloted aircraft systems into the {European} aviation system,''
  \emph{European RPAS Steering Group, Tech. Rep}, 2013.

\bibitem{8288376}
W.~{Khawaja}, O.~{Ozdemir}, and I.~{Guvenc}, ``{UAV} air-to-ground channel
  characterization for {mmWave} systems,'' in \emph{2017 IEEE 86th Vehicular
  Technology Conference (VTC-Fall)}, 2017, pp. 1--5.

\bibitem{8244753}
E.~{Greenberg} and P.~{Levy}, ``Channel characteristics of {UAV} to ground
  links over multipath urban environments,'' in \emph{2017 IEEE International
  Conference on Microwaves, Antennas, Communications and Electronic Systems
  (COMCAS)}, 2017, pp. 1--4.

\bibitem{8674482}
C.~{Calvo-Ramírez}, Z.~{Cui}, C.~{Briso}, K.~{Guan}, and D.~W. {Matolak},
  ``{UAV} air-ground channel ray tracing simulation validation,'' in \emph{2018
  IEEE/CIC International Conference on Communications in China (ICCC
  Workshops)}, 2018, pp. 122--125.

\bibitem{9211744}
Z.~{Ma}, B.~{Ai}, R.~{He}, G.~{Wang}, Y.~{Niu}, M.~{Yang}, J.~{Wang}, Y.~{Li},
  and Z.~{Zhong}, ``Impact of {UAV} rotation on {MIMO} channel characterization
  for air-to-ground communication systems,'' \emph{IEEE Transactions on
  Vehicular Technology}, vol.~69, no.~11, pp. 12\,418--12\,431, 2020.

\bibitem{8594724}
Q.~{Zhu}, K.~{Jiang}, X.~{Chen}, W.~{Zhong}, and Y.~{Yang}, ``A novel {3D}
  non-stationary {UAV-MIMO} channel model and its statistical properties,''
  \emph{China Communications}, vol.~15, no.~12, pp. 147--158, 2018.

\bibitem{8937764}
Z.~{Ma}, B.~{Ai}, R.~{He}, G.~{Wang}, Y.~{Niu}, and Z.~{Zhong}, ``A wideband
  non-stationary air-to-air channel model for {UAV} communications,''
  \emph{IEEE Transactions on Vehicular Technology}, vol.~69, no.~2, pp.
  1214--1226, 2020.

\bibitem{8633886}
H.~{Jiang}, Z.~{Zhang}, and G.~{Gui}, ``Three-dimensional non-stationary
  wideband geometry-based {UAV} channel model for {A2G} communication
  environments,'' \emph{IEEE Access}, vol.~7, pp. 26\,116--26\,122, 2019.

\bibitem{ArtigoCaracteristicasCanleComunicacionsUAV_2017}
J.~Rodr\'iguez-{Pi\~neiro}, T.~Dom\'inguez-Bola{\~n}o, X.~Cai, Z.~Huang, and
  X.~Yin, ``Air-to-ground channel characterization for low-height {UAVs} in
  realistic network deployments,'' \emph{{IEEE} Trans. Antennas Propag.}, vol.
  Early Access, 2020.

\bibitem{8576578}
X.~{Cai}, J.~{Rodr\'iguez-Pi\~neiro}, X.~{Yin}, N.~{Wang}, B.~{Ai}, G.~F.
  {Pedersen}, and A.~P. {Yuste}, ``An empirical air-to-ground channel model
  based on passive measurements in {LTE},'' \emph{{IEEE} Trans. Veh. Technol.},
  vol.~68, no.~2, pp. 1140--1154, Feb 2019.

\bibitem{hourani2018modeling}
A.~{Al-Hourani} and K.~{Gomez}, ``Modeling cellular-to-{UAV} path-loss for
  suburban environments,'' \emph{IEEE Wireless Communications Letters}, vol.~7,
  no.~1, pp. 82--85, 2018.

\bibitem{amorim2017radio}
R.~Amorim, H.~Nguyen, P.~Mogensen, I.~Z. Kov\'acs, J.~Wigard, and T.~B.
  Sorensen, ``Radio channel modeling for {UAV} communication over cellular
  networks,'' \emph{IEEE Wireless Communications Letters}, vol.~6, no.~4, pp.
  514--517, Aug 2017.

\bibitem{khawaja2016uwb}
W.~Khawaja, I.~Guvenc, and D.~Matolak, ``{UWB} channel sounding and modeling
  for {UAV} air-to-ground propagation channels,'' in \emph{IEEE Global
  Communications Conference (GLOBECOM)}, 2016, pp. 1--7.

\bibitem{ArtigoCaracteristicasCanleComunicacionsUAV_ParteTraxectorias_2017}
J.~Rodr\'iguez-{Pi\~neiro}, Z.~Huang, X.~Cai, T.~Dom\'inguez-Bola{\~n}o, and
  X.~Yin, ``Geometry-based {MPC} tracking and modeling algorithm for
  time-varying {UAV} channels,'' \emph{{IEEE} Trans. Wireless Commun.}, vol.
  Early Access, 2020.

\bibitem{8787874}
C.~{Yan}, L.~{Fu}, J.~{Zhang}, and J.~{Wang}, ``A comprehensive survey on {UAV}
  communication channel modeling,'' \emph{IEEE Access}, vol.~7, pp.
  107\,769--107\,792, 2019.

\bibitem{7486380}
D.~W. Matolak and R.~Sun, ``Air-ground channels for {UAS}: Summary of
  measurements and models for {L-} and {C-bands},'' in \emph{2016 Integrated
  Communications Navigation and Surveillance (ICNS)}, April 2016, pp.
  8B2--1--8B2--11.

\bibitem{8928089}
Y.~{Wang}, R.~{Zhang}, B.~{Li}, X.~{Tang}, and D.~{Wang}, ``Angular spread
  analysis and modeling of {UAV} air-to-ground channels at 3.5\,{GHz},'' in
  \emph{International Conference on Wireless Communications and Signal
  Processing (WCSP)}, 2019, pp. 1--5.

\bibitem{izydorczyk2019angular}
T.~Izydorczyk, F.~M.~L. Tavares, G.~Berardinelli, M.~C. Bucur, and P.~E.
  Mogensen, ``Angular distribution of cellular signals for {UAVs} in urban and
  rural scenarios,'' in \emph{European Conference on Antenna and Propagation
  (EuCAP)}, 2019.

\bibitem{garcia2019essential}
A.~{Garcia-Rodriguez}, G.~{Geraci}, D.~{Lopez-Perez}, L.~G. {Giordano},
  M.~{Ding}, and E.~{Bjornson}, ``The essential guide to realizing
  {5G}-connected {UAVs} with massive {MIMO},'' \emph{IEEE Communications
  Magazine}, vol.~57, no.~12, pp. 84--90, 2019.

\bibitem{geraci2018understanding}
G.~{Geraci}, A.~{Garcia-Rodriguez}, L.~{Galati Giordano},
  D.~{L{\'o}pez-P{\'e}rez}, and E.~{Bj{\"o}rnson}, ``Understanding {UAV}
  cellular communications: From existing networks to massive {MIMO},''
  \emph{IEEE Access}, vol.~6, pp. 67\,853--67\,865, 2018.

\bibitem{3GPP38901}
``{Study on channel model for frequencies from 0.5 to 100 GHz},'' Tech. Rep.,
  3GPP TR 38.901 V16.1.0, Jan. 2020.

\bibitem{8894135}
T.~{Izydorczyk}, F.~M.~L. {Tavares}, G.~{Berardinelli}, and P.~{Mogensen}, ``A
  {USRP}-based multi-antenna testbed for reception of multi-site cellular
  signals,'' \emph{IEEE Access}, vol.~7, pp. 162\,723--162\,734, 2019.

\bibitem{7928479}
X.~{Cai}, A.~{Gonzalez-Plaza}, D.~{Alonso}, L.~{Zhang}, C.~B. {Rodr\'iguez},
  A.~P. {Yuste}, and X.~{Yin}, ``Low altitude {UAV} propagation channel
  modelling,'' in \emph{11th European Conference on Antennas and Propagation
  (EuCAP)}, 2017, pp. 1443--1447.

\bibitem{753729}
B.~H. {Fleury}, M.~{Tschudin}, R.~{Heddergott}, D.~{Dahlhaus}, and K.~{Ingeman
  Pedersen}, ``Channel parameter estimation in mobile radio environments using
  the {SAGE} algorithm,'' \emph{IEEE Journal on Sel. Areas in Comm.}, vol.~17,
  no.~3, pp. 434--450, 1999.

\bibitem{xuesong_tap}
X.~{Cai}, W.~{Fan}, X.~{Yin}, and G.~F. {Pedersen}, ``Trajectory-aided
  maximum-likelihood algorithm for channel parameter estimation in
  ultra-wideband large-scale arrays,'' \emph{IEEE Transactions on Antennas and
  Propagation}, pp. 1--1, 2020.

\bibitem{1100705}
H.~{Akaike}, ``A new look at the statistical model identification,''
  \emph{{IEEE} Trans. Autom. Control}, vol.~19, no.~6, 1974.

\bibitem{8412215}
Y.~{Ji}, W.~{Fan}, and G.~F. {Pedersen}, ``Channel characterization for
  wideband large-scale antenna systems based on a low-complexity maximum
  likelihood estimator,'' \emph{{IEEE} Trans. Wireless Commun.}, vol.~17,
  no.~9, pp. 6018--6028, 2018.

\bibitem{4109300}
N.~{Czink}, P.~{Cera}, J.~{Salo}, E.~{Bonek}, J.~{Nuutinen}, and J.~{Ylitalo},
  ``A framework for automatic clustering of parametric {MIMO} channel data
  including path powers,'' in \emph{IEEE Vehicular Technology Conference},
  2006, pp. 1--5.

\bibitem{4469589}
N.~{Czink}, R.~{Tian}, S.~{Wyne}, F.~{Tufvesson}, J.~{Nuutinen}, J.~{Ylitalo},
  E.~{Bonek}, and A.~F. {Molisch}, ``Tracking time-variant cluster parameters
  in {MIMO} channel measurements,'' in \emph{the second International
  Conference on Communications and Networking in China}, Aug 2007, pp.
  1147--1151.

\bibitem{8509190}
C.~{Huang}, R.~{He}, Z.~{Zhong}, B.~{Ai}, Y.~{Geng}, Z.~{Zhong}, Q.~{Li},
  K.~{Haneda}, and C.~{Oestges}, ``A power-angle-spectrum based clustering and
  tracking algorithm for time-varying radio channels,'' \emph{IEEE Transactions
  on Vehicular Technology}, vol.~68, no.~1, pp. 291--305, 2019.

\bibitem{8240983}
X.~Cai, B.~Peng, X.~Yin, and A.~P. Yuste, ``Hough-transform-based cluster
  identification and modeling for {V2V} channels based on measurements,''
  \emph{IEEE Transactions on Vehicular Technology}, vol.~67, no.~5, pp.
  3838--3852, May 2018.

\bibitem{book:275224}
J.~F. Trevor~Hastie, Robert~Tibshirani, \emph{The Elements of Statistical
  Learning: Data Mining, Inference, and Prediction, Second Edition (Springer
  Series in Statistics)}, 2nd~ed., ser. Springer Series in Statistics.\hskip
  1em plus 0.5em minus 0.4em\relax Springer, 2009.

\bibitem{8103059}
C.~{Ling}, X.~{Yin}, R.~{M\"{o}ller}, S.~{H\"{a}fner}, D.~{Dupleich},
  C.~{Schneider}, J.~{Luo}, H.~{Yan}, and R.~{Thom\"{a}}, ``Double-directional
  dual-polarimetric cluster-based characterization of 70-77 {GHz} indoor
  channels,'' \emph{{IEEE} Trans. Antennas Propag.}, vol.~66, no.~2, pp.
  857--870, Feb 2018.

\bibitem{6691924}
C.~{Gustafson}, K.~{Haneda}, S.~{Wyne}, and F.~{Tufvesson}, ``On mm-wave
  multipath clustering and channel modeling,'' \emph{IEEE Transactions on
  Antennas and Propagation}, vol.~62, no.~3, pp. 1445--1455, March 2014.

\bibitem{7331737}
S.~{Cheng}, M.~{Martinez-Ingles}, D.~P. {Gaillot}, J.~{Molina-Garcia-Pardo},
  M.~{Li\'enard}, and P.~{Degauque}, ``Performance of a novel automatic
  identification algorithm for the clustering of radio channel parameters,''
  \emph{IEEE Access}, vol.~3, pp. 2252--2259, 2015.

\bibitem{1577605}
N.~{Czink}, P.~{Cera}, J.~{Salo}, E.~{Bonek}, J.~. {Nuutinen}, and
  J.~{Ylitalo}, ``Improving clustering performance using multipath component
  distance,'' \emph{Electronics Letters}, vol.~42, no.~1, pp. 33--45, Jan 2006.

\bibitem{MCD}
M.~Steinbauer, H.~Ozcelik, H.~Hofstetter, C.~F. Mecklenbrauker, and E.~Bonek,
  ``How to quantify multipath separation,'' \emph{IEICE Transactions on
  Electronics}, vol. E85-C, no.~3, pp. 552--557, Mar 2002.

\bibitem{9115069}
X.~{Cai}, G.~{Zhang}, C.~{Zhang}, W.~{Fan}, J.~{Li}, and G.~F. {Pedersen},
  ``Dynamic channel modeling for indoor millimeter-wave propagation channels
  based on measurements,'' \emph{{IEEE} Trans. Commun.}, vol.~68, no.~9, pp.
  5878--5891, 2020.

\end{thebibliography}

\end{document}